\documentclass{JHEP3}
\usepackage{epsfig}
\usepackage{epstopdf}
\input{epsf.sty}
\usepackage{epsf}
\usepackage{amssymb}
\newcommand{\nn}{\nonumber \\}
\newcommand{\ls}{\mathrel{\raise1.16pt\hbox{$<$}\kern-7.0pt %  <
\lower3.06pt\hbox{{$\scriptstyle \sim$}}}}         %  ~
\newcommand{\gs}{\mathrel{\raise1.16pt\hbox{$>$}\kern-7.0pt %  >
\lower3.06pt\hbox{{$\scriptstyle \sim$}}}}         %  ~
\def\VEV#1{{\langle #1 \rangle}}
\long\def\comment#1{}

\def\fun#1#2{\lower3.6pt\vbox{\baselineskip0pt\lineskip.9pt
  \ialign{$\mathsurround=0pt#1\hfil##\hfil$\crcr#2\crcr\sim\crcr}}}

\def\ba{\begin{eqnarray}}
\def\ea{\end{eqnarray}}
\def\be{\begin{equation}}
\def\ee{\end{equation}}
\def\nn{\nonumber \\}
\def\vk{{\bf k}}

\def\vx{{\bf x}}

\def\dthreek#1{\int\frac{d^3{\bf k_{#1}}}{(2\pi)^3}}

\title{Effects of Scale-Dependent Non-Gaussianity on Cosmological Structures}
\author{ Marilena LoVerde$^1$, Amber Miller$^1$, Sarah Shandera$^1$, Licia Verde$^2$\\
$^1$ Institute of Strings, Cosmology and Astroparticle Physics \\
Physics Department, Columbia University, New York, NY 10027\\
\vskip 0.3cm 
$^2$ ICREA, Institut de Ci\`encies de l'Espai (ICE), IEEC-CSIC, Campus UAB,
F. de Ci\`encies, Torre C5 par-2,  Barcelona 08193, Spain
\vskip 0.3cm
Emails: marilena@phys.columbia.edu, amber@astro.columbia.edu, sarah@phys.columbia.edu, verde@ieec.uab.es}
\abstract{ 
The detection of primordial non-Gaussianity could provide a powerful means to test various inflationary scenarios. Although scale-invariant non-Gaussianity (often described by the $f_{NL}$ formalism) is currently best constrained by the CMB, single-field models with changing sound speed can have strongly scale-dependent non-Gaussianity. Such models could evade the CMB constraints but still have important effects at scales responsible for the formation of cosmological objects such as clusters and galaxies. We compute the effect of scale-dependent primordial non-Gaussianity on cluster number counts as a function of redshift, using a simple ansatz to model scale-dependent features. We forecast constraints on these models achievable with forthcoming data sets. We also examine consequences for the galaxy bispectrum. Our results are relevant for the Dirac-Born-Infeld model of brane inflation, where the scale-dependence of the non-Gaussianity is directly related to the geometry of the extra dimensions.
}

\begin{document}

\maketitle

\section{Introduction}
Recent Cosmic Microwave Background (CMB) measurements e.g. \cite{Spergel:2006hy} have tightened constraints on cosmological parameters, verifying the inflationary predictions of a flat universe, structure formation from primordial adiabatic super-horizon perturbations and nearly scale invariant perturbations with a slightly red spectrum, and have started ruling out specific inflationary models (e.g. $\lambda \phi^4$ model). However, the observables measured so far have limited power to distinguish between scenarios, as the efforts to reconstruct the inflationary potential demonstrate e.g. \cite{Easther:2002rw, Peiris:2006sj, Powell:2007gu}. There are at least two possible observables, accessible in the near future, that have the potential to rule out or support large classes of models: non-Gaussianity and primordial gravitational waves. Here we focus on the first possibility, although both may be related in an interesting way which we review briefly in \S\ref{scaledepsec}. Non-Gaussianity is particularly valuable because it probes details of the inflaton self-interactions during inflation and can distinguish properties of the inflaton Lagrangian that cannot be constrained by the power spectrum alone. In this work, we would like to emphasize observable and scale-dependent non-Gaussianity as a signature of (so far) non-standard inflationary physics and demonstrate that near-future observations on a range of scales can provide important constraints for scale-dependent scenarios.

The simplest single field, slow-roll inflation predicts nearly Gaussian initial fluctuations \cite{Guth:1982ec, Starobinsky:1982ee, Bardeen:1983qw, Falk:1992sf, Gangui:1993tt, Gangui:1994yr, Wang:1999vf, Gangui:1999vg} where the deviation from Gaussianity is unobservable \cite{Acquaviva:2002ud, Maldacena:2002vr}. Multi-field models may or may not produce interesting non-Gaussianity, depending on the model \cite{Battefeld:2006sz}. For a review of possibilities known through 2004, see for example \cite{Bartolo:2004if}. However, even single-field inflation may generate significant non-Gaussianity if the inflaton has a non-trivial kinetic term\footnote{Single field models which temporarily violate the slow-roll conditions may also give significant non-Gaussianity. The case of a step in the potential is analyzed in \cite{Chen:2006xj} and is easily distinguishable from the scenarios discussed here.}. Mukhanov and others \cite{Armendariz-Picon:1999rj, Garriga:1999vw} have investigated the case of a Lagrangian which is a general function of the inflaton and powers of its first derivative. These models are characterized by a sound speed $c_s$ different from $1$ during inflation, which changes the time of horizon exit for scalar modes ($c_sk=aH$) but not for tensor modes. (Here $c_s$ is dimensionless, normalized by the speed of light $c=1$.) The primordial bispectrum of these scenarios has been worked out in detail \cite{Seery:2005wm, Chen:2006nt}, with the important result that higher-order derivative terms introduce new and dominant contributions to the three point function. For small sound speed, these terms can generate non-Gaussianity that is observably large. In addition, the sound speed may change during inflation, leading to scale-dependent non-Gaussianity.

A form of non-Gaussianity of initial conditions that is widely used in the literature is \cite{Salopek:1990jq, Verde:1999ij, Komatsu:2001rj}:
\be
\zeta({\bf x})=\zeta_G({\bf x})+\frac{3}{5}f_{NL}\left[\zeta_G^2({\bf x})-\langle\zeta_G^2({\bf x}) \rangle\right]
\label{eq:NG1}
\ee
where $\zeta({\bf x})$ is the primordial curvature perturbation, $\zeta_G({\bf x})$ is a Gaussian random field and the degree of non-Gaussianity is parameterized by (constant) $f_{NL}$. Here a positive $f_{NL}$ leads to a positive skewness in the density perturbations\footnote{There are several different conventions in the literature for defining $f_{NL}$ through an equation like Eq.(\ref{eq:NG1}). Here we use the same convention as WMAP \cite{Komatsu:2003fd}, but see Appendix \ref{fNLconv} for a discussion and details of some of the conventions used by previous authors.}. For non-Gaussianity of this type, CMB data are expected to yield the strongest constraints on $f_{NL}$ \cite{Verde:1999ij, Verde:2000vr}. Current CMB data (WMAP3) already constrain $-36<f_{NL}< 100$ \cite{Creminelli:2006rz} and could potentially achieve $|f_{NL}|\sim $ few \cite{Komatsu:2003fd, Creminelli:2005hu, Spergel:2006hy, Chen:2006ew}. 

The three-point function of the Fourier transform of Eq.(\ref{eq:NG1}) has a particular dependence on the three momenta ($k_1$, $k_2$, $k_3$) that has been dubbed the \emph{local} shape \cite{Babich:2004gb}. Scenarios where non-Gaussianity is generated outside the horizon (including the curvaton scenario \cite{Lyth:2002my} and the variable decay width model \cite{Zaldarriaga:2003my}) have this shape. A characteristic of the \emph{local} shape is that the magnitude of the three-point function is largest when one of the momenta is much smaller than the others (the squeezed limit). For scenarios different from the local case, an effective parameter $f^{eff}_{NL}$ may be defined from the magnitude of the three-point function evaluated at $k_1=k_2=k_3$ (the equilateral limit). This has been used to indicate the amount of non-Gaussianity in various models and is the variable CMB experiments report constraints on \cite{Komatsu:2003fd}. However, possibly a better single number to compare between models is the skewness (defined in \S \ref{NGdist}), which integrates over all shape configurations of the three-point function in $k$-space. 

The dominant contribution in single field models with higher derivative terms has a very different shape in Fourier space: it is largest when all momenta are equal. A momentum-dependent estimator which has its maximum in the equilateral limit and can be efficiently compared to data has been used to constrain the magnitude of equilateral type non-Gaussianity from the CMB \cite{Creminelli:2006gc, Creminelli:2006rz}. We will call the $k$-space configuration dependence of the estimator the \emph{equilateral} shape and the corresponding effective parameter $f^{eq}_{NL}$. From the WMAP three year data, Creminelli et al. find $-256<f^{eq}_{NL}<332$ \cite{Creminelli:2006rz}. The equilateral shape differs by about 20$\%$ from the more general sound speed case, as we review below. More importantly, the non-Gaussianity can be scale-dependent, and in this case observational constraints assuming scale-independence should be revisited\footnote{The constraints on $f_{NL}$ from WMAP are derived assuming a scale independent $f_{NL}$. Here we assume that the WMAP bound is valid for $f_{NL}(\ell_{max})$ where $\ell_{max}=475$ is the maximum scale used in the analysis. If the data were reanalyzed permitting a scale-dependent non-Gaussianity it is likely that the constraints on $f_{NL}$ would be relaxed.}. In this paper, we will concentrate on non-Gaussianity that runs (that is, has scale-dependence). Running non-Gaussianity may be a natural phenomenological question (and in fact even the slow-roll non-Gaussianity runs if the spectral index does), but it also has significant theoretical motivation. As we will review, the best-studied string theory model that gives rise to large non-Gaussianity has a natural scale-dependence. If the non-Gaussianity is large enough to be observable, its scale-dependence should also be investigated.

\begin{figure}[t]
\begin{center}
\includegraphics[width=0.7\textwidth,angle=0]{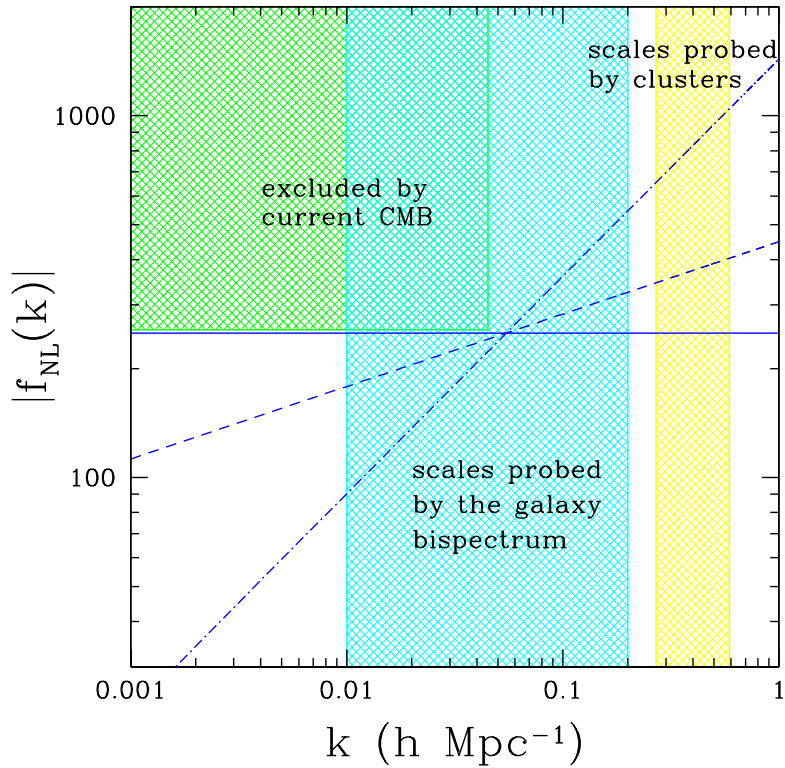} 
\caption{The quantity $|f_{NL}(k)|$ for several different values of the running of the non-Gaussianity $n_{NG}-1\equiv d\ln f_{NL}/d\ln k$. The solid line has $n_{NG}-1=0$, the dashed $n_{NG}-1=0.2$, and the dot-dashed $n_{NG}-1=0.6$. The shaded region in the upper left hand corner shows the range that is excluded at $95\%$ confidence by current CMB data \cite{Creminelli:2006rz} for \emph{equilateral} shape non-Gaussianity (plotted is the more conservative lower bound on $f_{NL}$) . The shaded regions on the right shows the range of scales probed by the galaxy bispectrum and by clusters. The range of scales probed by the bispectrum depends (among other things) on the redshift of the survey,  survey volume and the number density of galaxies, the above plot assumes $V\sim 10 h^{-3}Gpc^3$, $z\sim 1$ and the maximum $k$ is determined by the nonlinear scale \cite{Sefusatti:2007ih}.}

\label{fNLplot}
\end{center}
\end{figure}

Broadly speaking, constraints on primordial non-Gaussianity from observations of galaxies and clusters fall under two categories: observations of rare objects (such as galaxy clusters) which are sensitive to the tail of the probability distribution for density fluctuations, and measurements of higher-order clustering statistics of the density field. It has long been recognized that even small deviations from Gaussianity in the primordial fluctuation distribution would cause significant changes in the high mass tail of the halo distribution \cite{Lucchin:1987yv, Colafrancesco:1989px, Chiu:1997xb, Robinson:1998dx, Koyama:1999fc, Robinson:1999wh, Matarrese:2000iz, Verde:2000vr}. Therefore observations of rare and/or high-redshift collapsed objects could place interesting constraints on non-Gaussianity, particularly for models in which the magnitude of the non-Gaussianity increases significantly on small scales.  Observations of galaxies and clusters of galaxies provide valuable information on scales different from CMB observations, as illustrated in Figure \ref{fNLplot}. In addition, they are subject to a very different set of systematic errors and observational constraints, strengthening their utility as a complement to CMB anisotropy observations. 

Higher-order clustering statistics, such as the galaxy bispectrum, depend upon the correlation statistics of the primordial curvature perturbations \cite{Fry:1993xy,Chodorowski:1995qj,Scoccimarro:2000sn,Durrer:2000gi,Scoccimarro:2003wn,Hikage:2006fe,Sefusatti:2007ih}. Unlike tests of non-Gaussianity from the abundance of collapsed objects, the galaxy bispectrum retains information about the shape-dependence of the primordial three-point function. As pointed out in \cite{Sefusatti:2007ih} the bispectrum is potentially a powerful tool for distinguishing between the \emph{local} and \emph{equilateral} models we discuss in \S \ref{compareModels}. Measurements of the galaxy bispectrum can constrain $f_{NL}$ on scales of  tens of $h^{-1} Mpc$.  Current constraints from large scale structure, assuming a scale-independent $f_{NL}$ of the \emph{local} type, are at the level $|f_{NL}|\sim2000$ \cite{Feldman:2000vk} but future data set could achieve $|\Delta f^{local}_{NL}|\sim10$ and $|\Delta f^{equil}_{NL}|\sim100$ \cite{Scoccimarro:2003wn,Sefusatti:2007ih}.   In \S \ref{bispectrum} we briefly discuss the effects of scale-dependent primordial non-Gaussianity on the evolved bispectrum. 

Before beginning a detailed discussion of observables, it is interesting to note that from an effective field theory point of view one does not expect to see $f_{NL}$ more than a few without significant fine-tuning. For slow-roll models with standard kinetic terms, $f_{NL}$ is proportional to slow-roll parameters. For example, the contribution from the third derivative of the potential goes like $\frac{V^{\prime\prime\prime}}{2\pi H}\ll10^{-4}$, where the small number comes from the amplitude of scalar fluctuations and the inequality comes from demanding a flat enough potential for sufficient inflation. This was shown in detail by Maldacena \cite{Maldacena:2002vr}. Terms with a shift symmetry in the inflaton field $\phi$, as $\phi\rightarrow\phi+c$ (with $c$ a constant) may be added without spoiling slow-roll, and in particular terms like $(\nabla\phi)^{2n}/M^{4n-4}$, with $n>2$ and $M$ some mass scale, may be added. In order to truncate the added terms at some finite $n$ without fine-tuning the coefficients, we need $\nabla\phi<<M^2$, and so as shown by Creminelli \cite{Creminelli:2003iq} expect $f^{eff}_{NL}$ of order 1 at most.

However, string theory is a natural place to look for a fundamental description of inflation, and, as a UV complete theory, contains exceptions to this expectation. While there are many string-inspired scalar fields that have been explored for cosmology, brane inflation \cite{Dvali:1998pa} is an interesting example where a non-trivial kinetic term arises naturally. In this model the position of a brane (extended in our large dimensions and moving in the six compact extra dimensions) is the inflaton. The brane dynamics are given by the Dirac-Born-Infeld (DBI) action, where the kinetic term for the inflaton $\phi$ is proportional to $\sqrt{1-\dot{\phi}^2T^{-1}(\phi)}$, and the quantity $T$ is background dependent\footnote{There are certainly corrections to this action in a variety of contexts, but the crucial feature of the appearance of an exact summation of powers of derivatives should remain.}. Here the square root is a natural higher-dimension extension of the relativistic action for a point particle. The square root gives a speed limit for the brane ($\dot{\phi}^2<T(\phi)$), and the closer the brane velocity is to that limit, the smaller the sound speed, $c_s$, of the inflaton. The magnitude of non-Gaussianity goes as $1/c_s^2$, which is reasonable since a small sound speed means that many terms in the expansion of the square root are important. In the notation of \cite{Creminelli:2003iq}, as $c_s\rightarrow0$, $\nabla\phi/M^2\rightarrow1$ from below. In a realistic scenario the speed limit is related to the background geometry of the extra dimensions and so can be quite small and $\phi$ (time) dependent. The sound speed may change considerably during inflation, leading to scale-dependent non-Gaussianity. Although some work is still needed to see if DBI inflation can be embedded in a consistent compactification (see, e.g. \cite{McAllister:2007bg}), the appearance of the square-root kinetic term that gives the interesting features is quite suggestive of possible stringy signatures. Its importance is enhanced by the typical difficulty of finding flat potentials in string theory. Inflationary toy models with tachyons contain similar exact summations of a series of higher order operators \cite{Barnaby:2007yb}. 

From here, \S\ref{generalAction} reviews the relevant features of general single-field models, emphasizing scale-dependence and the consequent relations between observables. We use brane inflation to illustrate how scale-dependence may arise naturally in scenarios where the inflaton field sees a higher dimensional background. For readers wishing to avoid wading through equations, the key expression is Eq.(\ref{kdepend}) and its application in Eq.(\ref{runfNL}), which gives our ansatz for the scale-dependence of the non-Gaussianity. In \S\ref{compareModels} we compare the three-point functions for the local ($f_{NL}$), general sound speed, and equilateral models. Next, we derive in \S\ref{NGdist} and \S\ref{NGMF} a non-Gaussian probability distribution and mass function and \S\ref{predictions} presents current constraints and forecasts.  In \S\ref{bispectrum} we illustrate the effect of scale-dependence on the bispectrum. We summarize the utility of these observations and some issues relevant for future work in \S\ref{Conclusions}. Appendix \ref{PerturbSign} contains a clarification of previous conventions for $f_{NL}$. Some issues in deriving a useful non-Gaussian probability distribution function (which are also relevant for simulations) can be found in Appendix \ref{Distributions}. The second half of that appendix compares our mass function with a suggestion found previously in the literature.

\section{General Single-field Inflation: Review and Example}
\label{generalAction}
In this section we review general single-field models. For readers interested primarily in observational results, only the first subsection and Eq.(\ref{kdepend}) are necessary to continue to \S 3. In \S\ref{scaledepsec} we discuss relationships between observables when the sound speed is small and changing, with the relationship between non-Gaussianity and the tensor-scalar ratio perhaps the most interesting. The last subsection introduces an example from string theory that is easy to visualize, and where a changing sound speed is natural.

\subsection{General Single Field Formalism}
Here we follow the notation of Garriga and Mukhanov \cite{Garriga:1999vw}. The most general single-field model can be described by the action for the inflaton field $\phi$, including gravity:
\be
\label{genaction}
S=\frac{1}{2}\int d^4x \sqrt{-g}\:[M_p^2R+2P(X,\phi)]\;,
\ee
where $X=-\frac{1}{2}g^{\mu\nu}\partial_{\mu}\phi\partial_{\nu}\phi$ and $R$ is the curvature calculated from the 4D metric $g_{\mu\nu}$. The function $P$ is the pressure and the energy density $E$ is given by
\be
E=2XP,_{X}-P\;,
\ee
where $P,_{X}$ is the derivative of the pressure with respect to $X$. We use the Friedmann-Robertson-Walker metric with scale factor $a(t)$
\be
ds_4^2=-dt^2+a^2(t)\delta_{ij}dx^idx^j\;.
\ee
The sound speed $c_s$ is given by 
\be
c_s^2=\frac{P,_{X}}{E,_{X}}=\frac{P,_{X}}{P,_{X}+2XP,_{XX}}\;.
\ee
The Friedmann equation relates the Hubble parameter ($H=\frac{\dot{a}}{a}$) to the inflationary energy density as usual
\be
3M_p^2H^2=E\;.
\ee
Throughout we use the reduced Planck mass, $M_p=(8\pi G_N)^{-1/2}$. To have the usual equation of state, we assume the potential energy is the dominant contribution to the total energy. Two combinations of derivatives besides the sound speed appear in the 3-point function:
\ba
\Sigma&=&XP_{,X}+2XP_{,XX}\\\nonumber
\lambda&=&X^2P_{,XX}+\frac{2}{3}X^3P_{,XXX}\;.
\ea

These models are not necessarily slow-roll in the sense that $V,_{\phi\phi}/V$ may not be small. However, there are suitable parameters analogous to the usual slow-roll quantities that are useful for expressing observables. These make use of the Hubble parameter $H$ and its time derivatives instead of the potential (the Hamilton-Jacobi formalism). In the notation of \cite{Seery:2005wm, Chen:2006nt} these are
\ba
\label{params}
\epsilon&=&-\frac{\dot{H}}{H^2}\\\nonumber
\eta&=&\frac{\dot{\epsilon}}{H\epsilon}\\\nonumber
\kappa&=&\frac{\dot{c}_s}{Hc_s}\;.
\ea

Inflation takes place for $\epsilon <1$. When $\epsilon$, $\eta$ and $\kappa$ are all small, to first order the scalar index and the tensor/scalar ratio are given by
\ba
\label{nsr}
n_s-1&\approx&-2\epsilon-\eta-\kappa\\\nonumber
r&=&16\epsilon c_s\;.
\ea
The power spectrum of the primordial curvature perturbation is given by
\be
\langle \zeta(\vk_1)\zeta (\vk_2)\rangle = (2\pi)^3\delta_D(\vk_1+\vk_2) \frac{(2\pi)^3\mathcal{P}^\zeta(k_1)}{4\pi k_1^3}
\label{pdef}
\ee
where $\mathcal{P}^\zeta(k)\propto k^{n_s-1}$ is the dimensionless variance and $\delta_D$ is the Dirac delta function. 

A detailed calculation of the bispectrum for models with arbitrary sound speed can be found in \cite{Chen:2006nt}. The non-Gaussianity is enhanced by $1/c_s^2$, and new terms are generically present that are not suppressed by the small parameters $\epsilon$ and $\eta$. These terms vanish when $c_s=1$ (equivalently $P,_{XX}=0$). Notice that in the slow-roll case, self-couplings of the inflaton that appear in the potential must be small to preserve slow-roll and the small amplitude of fluctuations. For small sound speed models, higher-order derivative terms do not interfere with slow-roll and lead to terms in the 3-point function which are not suppressed by $\epsilon$, $\eta$.

For the general action in Eq.(\ref{genaction}), the leading order contributions to the three point function of the primordial curvature may be written  \cite{Chen:2006nt}
\be
\VEV{\zeta(\vk_1)\zeta(\vk_2)\zeta(\vk_3)}=(2\pi)^7\delta_D(\vk_1+\vk_2+\vk_3)\frac{\mathcal{P}^\zeta(K)^2}{k_1^3k_2^3k_3^3}\left(\mathcal{A}_{\lambda}+\mathcal{A}_{c}+\mathcal{O}(\epsilon/c_s^2)+\mathcal{O}(\epsilon)\right)
\label{SS3pt}
\ee
where $K=k_1+k_2+k_3$ and 
\ba
\label{Adefs}
\mathcal{A}_{\lambda}&=&\left(\frac{1}{c_s^2}-1-\frac{\lambda}{\Sigma}[2-(3-2{\bf c_1})l]\right)_K\frac{3k_1^2k_2^2k_3^2}{2K^3}\\\nonumber
\mathcal{A}_{c}&=&\left(\frac{1}{c_s^2}-1\right)_K\left(-\frac{1}{K}\sum_{i>j}k_i^2k_j^2+\frac{1}{2K^2}\sum_{i\neq j}k_i^2k_j^3+\frac{1}{8}\sum_{i}k_i^3\right)\\\nonumber
\ea
where ${\bf c_1}\approx0.577$ is the Euler constant, and $l=\dot{\lambda}/(H\lambda)$ is of the same order as the slow-roll parameters\footnote{Notice that terms are not grouped completely by slow-roll order, so that some terms in $\mathcal{A}_{\lambda}$ are of the same order as terms we have not explicitly written out.}. There are several terms suppressed by slow-roll parameters hidden in the $\mathcal{O}(\epsilon/c_s^2)$ and $\mathcal{O}(\epsilon)$ terms in Eq.(\ref{SS3pt}). These should also be evaluated to give a precise answer for any model. However, here we are concerned with an investigation of the magnitude of scale-dependence that can be constrained by experiment so we ignore these terms. Notice that the dimensionless power spectrum $\mathcal{P}^{\zeta}(K)$, evaluated at $K=k_1+k_2+k_3$ has been pulled out. This is a choice (a similar one is often made in the $f_{NL}$ formalism), but the full $k$-dependence of Eq.(\ref{SS3pt}) is of course fixed and can be written independently of the power spectrum.

The sign of the dominant piece depends on details of the Lagrangian. In particular the first line above, $\mathcal{A}_{\lambda}$, is not suppressed by a slow-roll parameter and can a priori have either sign. On the other hand since $c_s<1$ the second term,  $\mathcal{A}_c$,  is negative. For the DBI model presented in the next section, $2\lambda/\Sigma=1/c_s^2-1$ so the largest terms in $\mathcal{A}_{\lambda}$ vanish,  $\mathcal{A}_c$ dominates and the skewness is negative. The consequences that we show for the $\mathcal{A}_c$ term (which we call the ``$\mathcal{A}_c$ shape") are roughly those of the DBI model.

\subsection{Scale-Dependent Relationships}
\label{scaledepsec}
Scale dependence in the sound speed means that there is correlated scale dependence among the observables. In order for Eq.(\ref{nsr}) to be valid, we assume the parameter $\kappa$ is a good expansion parameter ($|\kappa|<1$). The sound speed may vary more quickly, but then the analysis of even the two-point function must be revisited (see discussion in \cite{Chen:2006nt}). When the sound speed $c_s$ does not change too rapidly, its scale dependence can be characterized in a manner similar to that of power spectrum \cite{Chen:2005fe}. That is, 
\be
\label{kdepend}
n_{NG}-1\equiv\frac{d\ln(c_s^{-2})}{d\ln\;k}\;.
\ee
Notice that for $\kappa<0$ the amount of non-Gaussianity increases with scale. For models where $\dot{\phi}=-2M_p^2H^{\prime}c_s$ we have
\be
\label{nNGkap}
n_{NG}-1\sim-2\kappa\;.
\ee

We saw in Eq.(\ref{nsr}) that the usual slow-roll consistency equation for the tensor/scalar ratio is modified even for a constant sound speed $c_s\neq1$.
\be
r=16\epsilon c_s.
\ee
This expression demonstrates an interesting relationship between measurable values of $r$ and measurable non-Gaussianity. Clearly, a very small sound speed will depress the ratio. However, from effective field theory with a constant sound speed, we do not expect to find observable tensor modes anyway. This expectation is commonly expressed as the Lyth bound \cite{Lyth:1996im}
\be
\frac{1}{M_p}\frac{d\phi}{dN_e}=\sqrt{\frac{r}{8}}\;.
\ee 
Which, for $r$ nearly constant, says that requiring $\Delta\phi/M_p<1$ and about 60 e-folds suggests $r$ less than a percent. The CMB plus SDSS results constrain $r<0.28$ at $k=0.002Mpc^{-1}$ \cite{Spergel:2006hy}, which is far above this expectation. If the sound speed changes during inflation then $r$ is no longer constant and it may be possible to find a somewhat larger tensor/scalar ratio at CMB scales. We would then also expect a non-Gaussian signal which is smaller on the largest scales and increasing at smaller scales. 

\subsection{An Example from String Theory}
\label{DBIexample}
Here we introduce brane inflation as a model for generating and constraining large non-Gaussianity. We review a few pertinent details and discuss why it is such an informative context in which to consider smaller scale limits on non-Gaussianity. We introduce it purely as a case in point, demonstrating that small and changing sound speed arises naturally in string theory.

From a model-building perspective, brane inflation has several attractive features. First, it sits in the well-explored arena of IIB flux compactifications \cite{Kachru:2003aw, Kachru:2003sx} and has been extensively compared with observations \cite{Firouzjahi:2005dh, Shandera:2006ax, Bean:2007hc, Peiris:2007gz}. More importantly, it can potentially exhibit stringy signatures, like substantial non-Gaussianity. Many models of inflation in string theory have inflaton potentials that are too steep to naturally generate 55-60 e-folds. In brane inflation the potential is also expected to be rather steep \cite{Kachru:2003sx, Krause:2007jk, Baumann:2007ah}, but the square-root kinetic term allows enough inflation by forcing the brane to ``slow-roll" in the sense of small $\dot{\phi}$ \cite{Silverstein:2003hf, Alishahiha:2004eh}. The field then spends long enough rolling down to generate enough inflation. Many of the interesting observational features (including non-Gaussianity) are linked to the ``speed limit" feature that is so useful for obtaining inflation. However, it is not yet clear if there is an honest way to embed this model in a fully consistent compactification. Experiment may resolve the issue faster than theory, as many parts of the parameter space can be ruled out in the near future as constraints on the tensor/scalar ratio and non-Gaussianity are improved.

The original brane inflation model used the attractive force between a $D$3-brane and an anti-brane ($\overline{D3}$) as the inflaton potential, with the brane separation as the inflaton. This simple idea can be made more realistic by placing the branes in a compactified description of the extra dimensions. We will not worry about the details of such a procedure beyond assuming a reasonable local background geometry that gives rise to scale-dependent non-Gaussianity. It is a property of that background (the warping) that causes the sound speed to change during inflation.

In brane inflation the canonical inflaton $\phi$ is related to the brane position. The basic action for a $D$-brane is the Dirac-Born-Infeld (DBI) action together with a potential $V(\phi)$ that is in principle known. Writing this action in a simplified form gives
\be
S=-\int d^4x\;a^3(t)\left[T(\phi)\sqrt{1- \dot{\phi}^2/T(\phi)} + V(\phi)\right]
\label{DBIact}
\ee
where $a(t)$ is the usual scale factor and $T(\phi)$ is a non-trivial function of $\phi$ that comes from the background six-dimensional geometry the brane moves in. The importance of the square root for the inflationary properties has led to the name `DBI model' for brane inflation in a $\phi$-dependent background geometry. The effects of the square root are usefully captured by a Lorentz factor $\gamma$:
\be
\label{gamma1}
\gamma(\phi)=\frac{1}{\sqrt{1-\dot{\phi}^2T^{-1}}}
\ee
which again demonstrates that the speed limit for $\dot{\phi}$ depends on the function $T(\phi)$. A functional form that arises naturally is $T(\phi)\sim\phi^4$ with corrections so that $T(\phi)$ goes to a constant as $\phi$ goes to 0 \cite{Klebanov:2000hb}. Then $\gamma$ may grow quite large at small $\phi$. In the case of the above square root action, there is a simple relationship between $\gamma$ and the sound speed:
\be
c_{s, DBI}=\frac{1}{\gamma}
\ee
For a brane moving toward small $\phi$ in a geometry that gives $T(\phi)\sim\phi^4$, the speed limit ($\dot{\phi}^2<T(\phi)$) becomes more constraining, so $\gamma$ grows during inflation and the non-Gaussian signature (which is $\propto 1/c_s^2-1$) increases. The $\phi$-dependence of the non-Gaussianity becomes scale-dependence in the observed primordial gravitational field as the fluctuations are frozen in at horizon crossing, $c_sk=aH$. The parameter $\kappa$ captures the change in sound speed, and in this paper we will treat $\kappa$ as a constant (although in the full solution it is not, and changes in $\kappa$ can distinguish features of the geometry). In the limit of large $\gamma$, $\kappa\rightarrow-2\epsilon$ and $\eta\rightarrow0$, so $n_s\rightarrow1$. In a variation of this scenario, the brane may start in the bottom of the throat and move out relativistically  \cite{Chen:2005fe, Chen:2004gc, Chen:2005ad, Bean:2007eh}. In this case the speed limit constraint decreases as inflation progresses, so the non-Gaussianity would be largest at larger scales. In other words, the scale dependence of the non-Gaussianity provides information about the geometry of the extra dimensions.

\section{Sizes, Shapes and Scalings}
So far we have discussed scale-dependence of non-Gaussianity which can arise in models of inflation with varying speed of sound. An equally important feature is the configuration dependence of the higher-order correlation functions. In this section we review and discuss the value of two categories of configuration dependence or \emph{shape} of the primordial bispectrum and how these may be adapted to allow for scale dependence. 
 
\label{compareModels}
\subsection{The Importance of Shape}
As discussed in \cite{Babich:2004gb}, two qualities distinguish the primordial bispectrum from sound speed models from that of local models. The first is {\it shape}, or the difference in the dependence of the three-point functions on the relative magnitudes of $k_1$, $k_2$ and $k_3$. The second is the scale dependence of the sound speed $c_s(k)$ which gives rise to a scale-dependent non-Gaussianity which we will address in the next section.

In the literature the non-Gaussianity is often assumed to be of the ``local" type defined through Eq.(\ref{eq:NG1}). The amplitude of $\VEV{\zeta_G^2}$ is expected to be small so that the expansion is a good approximation so long as $|f_{NL}| << \langle \zeta_G^2\rangle^{-1/2}$. Measurable quantities such as correlation functions of the non-Gaussian field $\zeta$ can be worked out from Eq.(\ref{eq:NG1}), 
\ba
\label{fnl3pt}
\VEV{\zeta(\vk_1)\zeta(\vk_2)} &=&\VEV{\zeta_G({\vk_1})\zeta_G(\vk_2)}+\mathcal{O}(f_{NL}^2\VEV{\zeta_G^2}) \\
\VEV{\zeta(\vk_1)\zeta(\vk_2)\zeta(\vk_3)}&=&(2\pi)^7\delta_D(\vk_1+\vk_2+\vk_3)\frac{\mathcal{P}^\zeta(K)^2}{k_1^3k_2^3k_3^3}\mathcal{A}_{local}(k_1,k_2,k_3)+\mathcal{O}(f_{NL}^3\VEV{\zeta_G^2}^{3/2})\nonumber
\ea
where \footnote{Here and throughout this paper we assume the variance, $\mathcal{P}^\zeta$ is described by a power law, that is $n_s-1$ is independent of $k$. Equations allowing for a general $\mathcal{P}^\zeta$ would be more complicated.}
\be
\mathcal{A}_{local}(k_1,k_2,k_3)=\frac{3f_{NL}}{10}K^{-2(n_s-1)}\left(k_1^3(k_2k_3)^{n_s-1}+k_2^3(k_1k_3)^{n_s-1}+k_3^3(k_1k_2)^{n_s-1}\right).
\ee

\begin{figure}[h]
\begin{center}
$\begin{array}{cc}
\includegraphics[width=0.5\textwidth,angle=0]{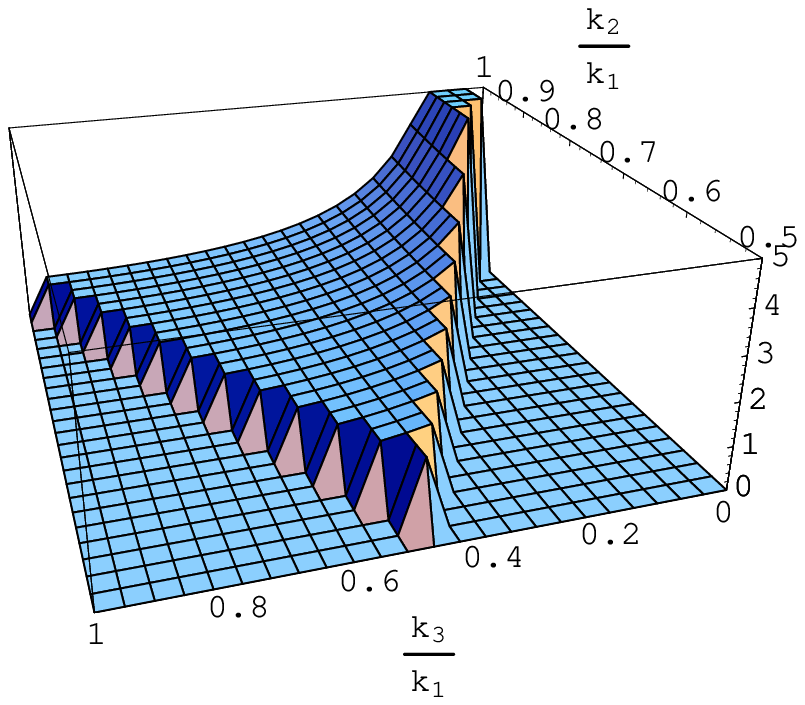} &
\includegraphics[width=0.4\textwidth,angle=0]{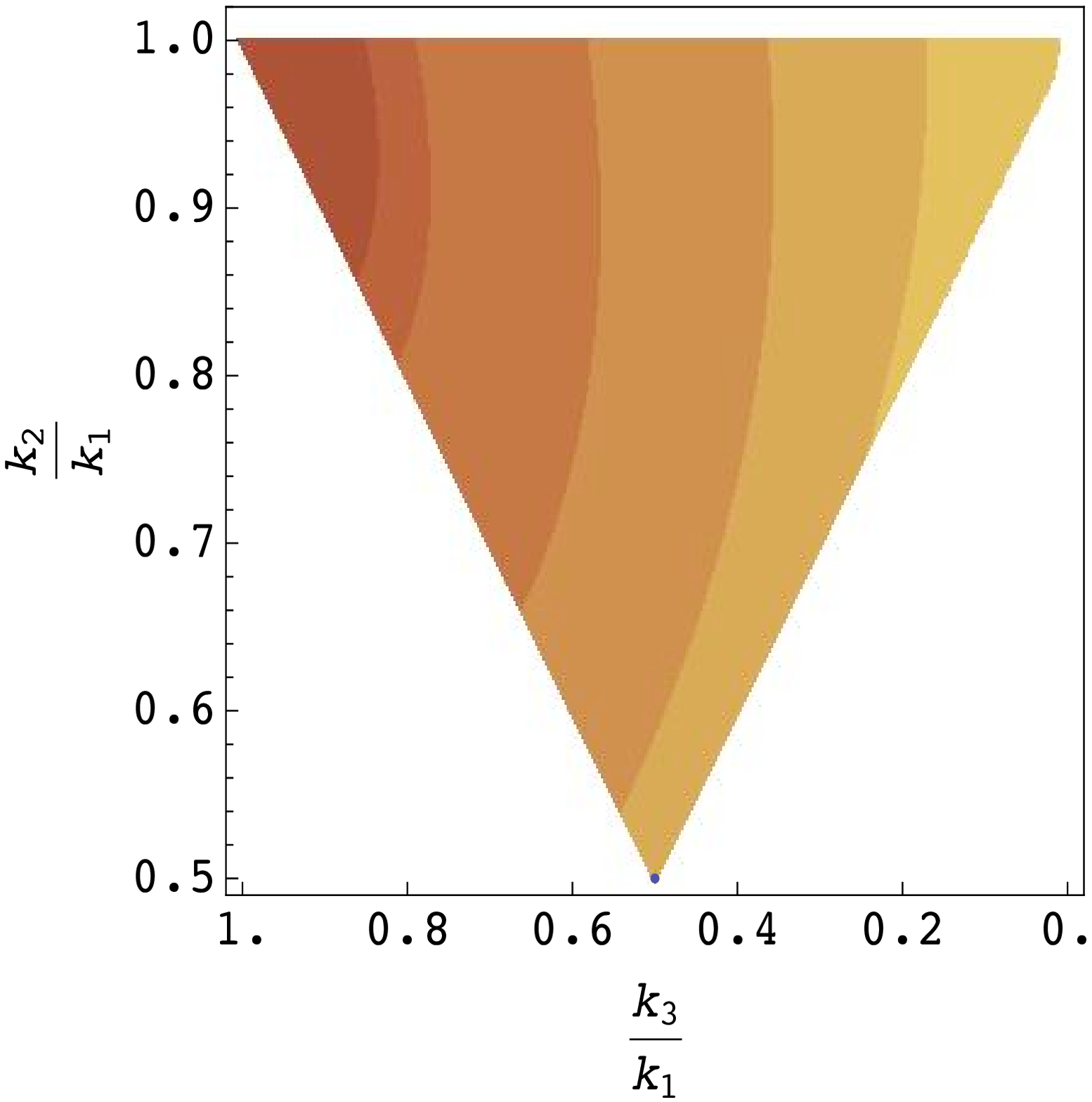} \\
\mbox{(a)}&\mbox{(b)}\\
\includegraphics[width=0.5\textwidth,angle=0]{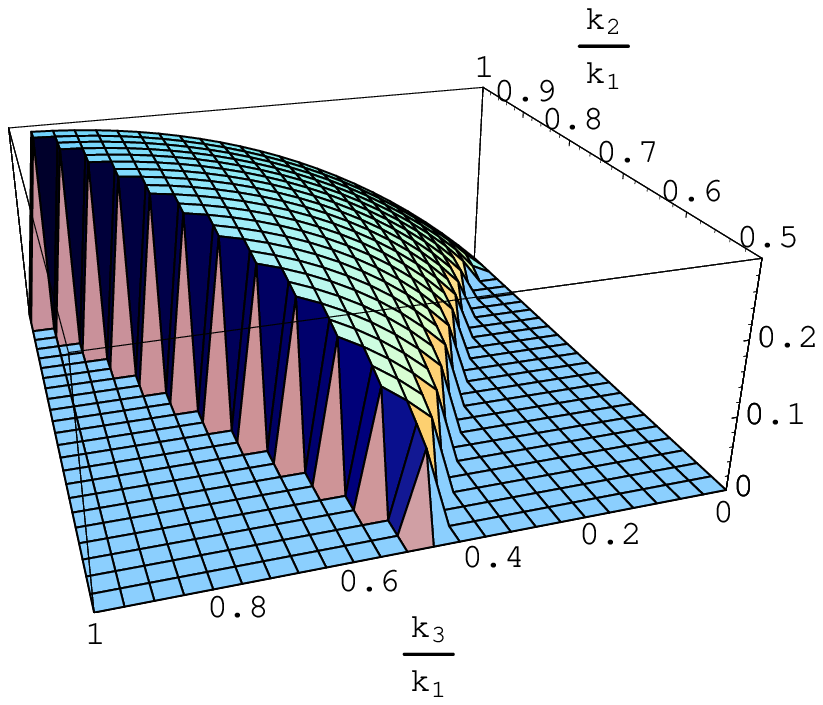} &
\includegraphics[width=0.4\textwidth,angle=0]{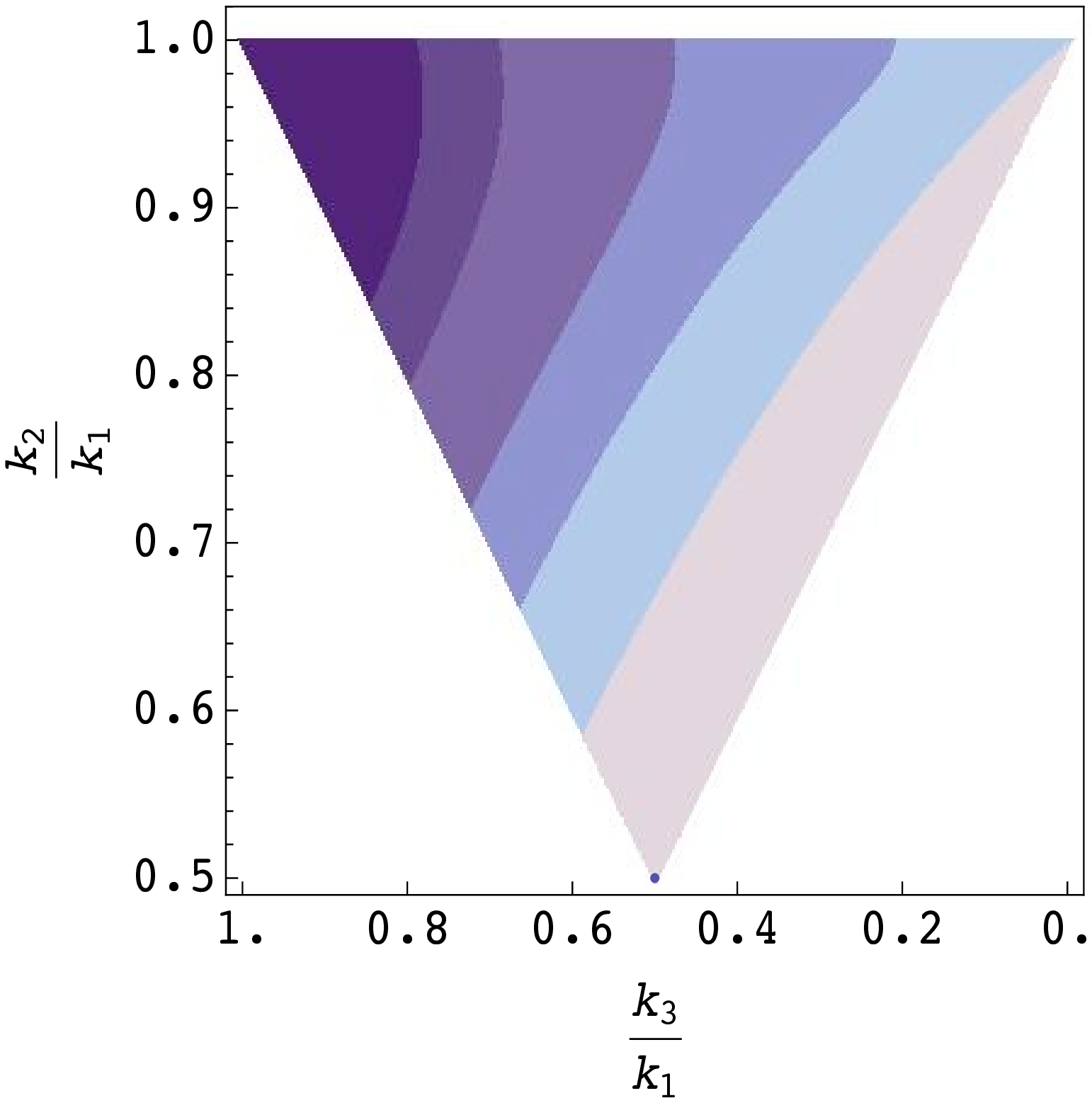} \\
\mbox{(c)} & \mbox{(d)}
\end{array}$
\caption{ (a) The shape of the primordial bispectrum for the local model, $\mathcal{A}_{local}(1,k_2,k_3)/(k_2k_3)/f_{NL}$.  The domain of the plot is restricted to $\vk_1+\vk_2+\vk_3=0$. (b) Contour plot of the fractional difference between the local form of non-Gaussianity and the DBI shape. Shaded regions show contours of (beginning from the upper left-hand corner) $(\mathcal{A}_{local}-\mathcal{A}_c)/\mathcal{A}_c=$$0$, $0.05$, $0.1$, $0. 5$, $1$, $2$, $10$. (c) The dominant shape in the primordial bispectrum for the DBI model, plotted is $\mathcal{A}_c(1,k_2,k_3)/(k_2k_3)/f_{NL}^c$.(d) Contour plot of the fractional difference between the equilateral form of non-Gaussianity and the DBI shape. Shaded regions show contours of (beginning from the upper left-hand corner) $(\mathcal{A}_{equil}-\mathcal{A}_c)/\mathcal{A}_c=$ $0$, $0.01$, $0.02$, $0.05$, $0. 1$, $0.25$. }
\label{Afigs}
\end{center}
\end{figure}

For general non-Gaussianity the 3-point function will not have the above form. In particular if we compare Eq.(\ref{fnl3pt}) and Eq.(\ref{SS3pt}) we see that the $k$ dependence is quite different, this is illustrated in panels (a) and (b) of Figure \ref{Afigs}. One can nevertheless define an $f^{eff}_{NL}$ by taking the equilateral triangle limit ($k_1=k_2=k_3$) of Eq.(\ref{SS3pt}) and Eq.(\ref{fnl3pt}), and identifying the coefficients\footnote{Note that our definitions of $f_{NL}$ for sounds speed models differ from those in CHKS \cite{Chen:2006nt}, see Appendix \ref{PerturbSign} for a discussion of this issue.}:
\ba
\label{fnls}
f^{eff}_{NL}&=&f_{NL}^\lambda+f_{NL}^c +\mathcal{O}(\epsilon/c_s^2)+\mathcal{O}(\epsilon)\nn
f_{NL}^{\lambda}&=&\frac{5}{81}\,3^{2(n_s-1)}\left(\frac{1}{c_s^2}-1-\frac{\lambda}{\Sigma}[2-(3-2c_1)l]\right)\nn
f_{NL}^{c}&=&-\frac{35}{108}\,3^{2(n_s-1)}\left(\frac{1}{c_s^2}-1\right)\;.
\ea
Notice that Eq.(\ref{eq:NG1}) assumes a constant $f_{NL}$, while the models described in \S \ref{generalAction} generally have $c_s(k)$. The $k$-dependence causes additional difficulty with assuming the local parameterization for non-Gaussianity. 

One could imagine using the local form of the non-Gaussianity with the above definitions of $f^{eff}_{NL}$ to estimate the non-Gaussianity from sound speed models, but this approximation is not very good. This fact is illustrated in panel (b) of Figure \ref{Afigs}, where the fractional difference between $\mathcal{A}_c$ (the dominant contribution for DBI inflation) and the local form $\mathcal{A}_{local}$ using the $f^{eff}_{NL}$ value in Eq.(\ref{fnls}) is plotted as a function of $k_2/k_1$ and $k_3/k_1$. The fractional difference is $\gs 1$ for much of the allowed $k_2/k_1$-$k_3/k_1$ range. 

\subsection{A Scale-Dependent Equilateral Model}
Creminelli et al. \cite{Creminelli:2005hu,Creminelli:2006rz} have proposed a functional form which closely approximates the behavior of the three-point function when a single higher derivative term, $(\nabla\phi)^4/(8M^4)$, has been included in the Lagrangian and gravity has been ignored. This shape is also useful for efficient data analysis. This so-called equilateral shape has a three-point function with the form
\ba
\label{AfNL}
\mathcal{A}_{equil.}(k_1,k_2,k_3)&=&\frac{9}{10}f^{eq}_{NL}K^{-2(n_s-1)}\left(-k_1^3(k_2k_3)^{n_s-1}+\textrm{perm.}-2(k_1k_2k_3)^{1+2(n_s-1)/3}\right.\nn
&+&\left.k_1^{2+(n_s-1)/3}k_2^{1+2(n_s-1)/3}k_3^{n_s-1}+\textrm{perm.}\right)\;.
\ea
Unlike the local shape, the maximum signal in $\mathcal{A}_{equil}(k_1,k_2,k_3)/(k_1^3k_2^3k_3^3)$ occurs for configurations with $k_1\approx k_2\approx k_3$. Recall that in the local shape the maximum contributions to the skewness occur for ``squeezed" configurations $k_1,k_2 >>k_3$. Current (95\% confidence level) constraints on the equilateral shape from the CMB are \cite{Creminelli:2006rz}
 \be
 -256 < f^{eq}_{NL}< 332\;. 
 \label{fnleqbounds}
 \ee
 
The parameterization of Eq.(\ref{eq:NG1}) and previous work on the equilateral shape assume a scale independent $f_{NL}$; with the possibility of scale-dependent non-Gaussianity the bounds on an $f_{NL}$-type parameter must be determined at various scales. Unlike the local model, there is no requirement that $f^{eq}_{NL}$ for the equilateral model be constant\footnote{Of course, one can imagine a configuration that is maximal for squeezed configurations but is not the local type as given in Eq.(\ref{eq:NG1}). Then there can certainly be scale-dependence. This is precisely the situation with the terms from standard slow-roll: they are maximum in the squeezed limit and proportional to parameters that may change during inflation.}. As discussed in \S\ref{scaledepsec}, inflationary models with changing sound speed predict that the non-Gaussianity will run with scale. We can extend any model, including the equilateral model, to allow for scale-dependent non-Gaussianity by setting
 \be
 \label{runfNL}
f^{eff}_{NL}(k)=f^{eff}_{NL,\:CMB}\left(\frac{k}{k_{CMB}}\right)^{-2\kappa}\;.
 \ee
where $|\kappa|\ll1$ is a free parameter. This choice is motivated by Eq.(\ref{nNGkap}), but we only consider the simplest choice that $\kappa$ is constant, at least between CMB and cluster scales. The choice of pivot scale given as $k_{CMB}$ in Eq.(\ref{runfNL}) is of course arbitrary. In this paper we are primarily interested in constraining non-Gaussianity that increases at small scales (large $k$) so we choose the pivot scale to be just below the minimum scale that has been used to constrain non-Gaussianity in the CMB, $k_{CMB}=0.04 \, Mpc^{-1}$. In this case $\kappa<0$. For non-Gaussianity that decreases on small scales, $\kappa>0$.

In Figure \ref{Afigs} the shapes of $\mathcal{A}_{local}$ and $\mathcal{A}_{equil}$ are compared with the shape of $\mathcal{A}_c$. The equilateral model is clearly a superior approximation to the true $\mathcal{A}_c$ shape. However the two shapes still differ by a non-negligible amount so we will continue to show predictions for both the equilateral and the ``DBI-like'' $\mathcal{A}_c$ shape. Notice from Eq.(\ref{fnls}) that the DBI model gives a negatively skewed distribution for the density fluctuations. In the case of the DBI model (keeping only the $c$ term in Eq.(\ref{Adefs}) and ignoring the difference between the the equilateral model and the $c$ term) Eq.(\ref{fnleqbounds}) gives a constraint on the sound speed at $ k_{CMB}$
 \be
 c_{s,DBI}(k_{CMB})\gs 0.034\;. 
 \label{csfNLdef}
 \ee

\section{Tests of Non-Gaussianity on Sub-CMB Scales: Cluster Number Counts}
\label{clustersintro}
 If non-Gaussianity increases at small scales, then it could evade CMB constraints but still leave an observational imprint on large-scale structures. There are at least two ways of testing non-Gaussianity at small scales: through higher-order correlations of large-scale structure probes (e.g. galaxy bispectrum and the three-point correlation function) and, reaching even smaller scales, the abundance of collapsed objects.  The first approach relies on knowledge of higher-order correlations of the primordial curvature perturbation which are then evolved by means of cosmological perturbation theory (for a review see \cite{Bernardeau:2001qr}). The second approach requires knowledge of the probability distribution from which the smoothed density fluctuation is drawn. This probability distribution is most sensitive to the skewness of the primordial curvature perturbation (assuming reasonable ordering of the cumulants) but in principle requires knowledge of an infinite number higher-order correlation functions. The evolved bispectrum retains the full information about the shape of the primordial bispectrum which is lost in the smoothed skewness used for predicting the abundance of collapsed objects.  Both the galaxy bispectrum and the abundance of collapsed objects are sensitive to the scale-dependence of the primordial bispectrum. We focus on cluster number counts first since these probe smaller scales, the bispectrum is considered in \S \ref{bispectrum}.
 
While in the derivations we will try to be as general as possible we make a few assumptions about the cosmology. In what follows we assume a flat $\Lambda CDM$ universe. In plots and calculations that require input of cosmological parameters we assume the WMAP III maximum likelihood values \cite{Spergel:2006hy}: a Hubble parameter today of $H_0=100h$ with $h=0.73$, $\Omega_m=0.24$, $\Omega_\Lambda=0.76$ and $\Omega_bh^2=0.0223$ as the fractional densities in matter, vacuum and baryons today, a scalar spectral index $n_s=0.958$, and that our power spectrum is normalized to $\sigma_8=0.77$ (corresponding to $\mathcal{P}^\zeta(k)\approx 2.22\times 10^{-9}\left(\frac{k}{0.002 Mpc^{-1}}\right)^{n_s-1}$). We will use the transfer function of \cite{Bardeen:1985tr} with a modified shape parameter $\Gamma=\Omega_m h\, \textrm{exp}[-\Omega_b(1+\sqrt{2h}/\Omega_m)]$ to calculate the linear power spectrum for matter perturbations  \cite{Sugiyama:1994ed}. 

\subsection{The Non-Gaussian Probability Distribution Function for the Smoothed Density Fluctuation}
\label{NGdist}
We are interested in predictions for rare objects, that is the collapsed objects that form in extreme peaks of the density field $\delta({\bf x})=\delta\rho/\rho$. The statistics of collapsed objects can be described by the statistics of the density perturbation smoothed on some length scale $R$ (or equivalently a mass scale $M=\frac{4}{3}\pi R^3 \rho$). This is given by
\be
\delta_R(z)=\dthreek{}  W_R(k)\delta(\vk,z)
\ee
where $W_R(k)$ is the Fourier transform of a window function, which we take to be a top-hat in real space giving
\be
W_R(k)=\frac{3\sin(kR)}{k^3R^3}-\frac{3\cos(kR)}{k^2R^2}.
\ee
The relation between the primordial curvature perturbation $\zeta$ and the linear perturbation to the matter density $\delta =\delta \rho/\rho$ today is
\ba
\delta(\vk,z)&=&M(k,z)\zeta(\vk)\nn
M(k,z)&=&\frac{2}{5}\frac{1}{\Omega_{m}}\frac{1}{H_0^2}D(z)T(k)k^2
\label{Mka}
\ea
where $D(z)$ is the linear growth function, $z$ is the redshift, and $T(k)$ is the transfer function which describes the suppression in amplitude of the modes that entered the horizon during the radiation dominated era. The linear matter power spectrum is then given by $P_{L}(k)=2\pi^2 M(k,z)^2\mathcal{P}^\zeta(k)/k^3$.

To incorporate non-Gaussian initial conditions into predictions for the smoothed density field we need an expression for the probability distribution function (PDF) for $\delta_R$. For a particular real-space expansion like Eq.(\ref{eq:NG1}), one may make a formal change of variable in the Gaussian PDF to generate a normalized distribution \cite{Matarrese:2000iz}. However, as we discuss below this expansion is not particularly well-matched to the DBI scenario, and it is not obvious what expression analogous to Eq.(\ref{eq:NG1}) captures the full $k$-dependence of realistic models. However, there is a straightforward mathematical relationship which allows one to build up the PDF from the cumulants and a known distribution like the Gaussian. This technique leads to the well-known Edgeworth expansion and is especially useful for quantities that depend on the central part of the PDF. We will demonstrate that it is also valid in a regime that allows us to probe clusters, at least for small enough masses and redshifts. 

For a probability density function $P(\delta_R)d\delta_R$, the $n$-th central moment is
\be
\langle\delta_R^n\rangle\equiv\int_{-\infty}^{\infty}\delta_R^nP(\delta_R)d\delta_R\;.
\ee
The $n$-th cumulant $\kappa_n$ is the connected $n$-point function, so that
\ba
\kappa_1&=&\langle\delta_R\rangle_c\\\nonumber
\kappa_2&=&\langle\delta_R^2\rangle-\langle\delta_R\rangle_c^2\\\nonumber
\kappa_3&=&\langle\delta_R^3\rangle - 3\langle\delta_R^2\rangle_c\langle\delta_R\rangle_c - \langle\delta_R\rangle_c^3\\\nonumber
\kappa_4&=&\langle\delta_R^4\rangle-4\langle\delta_R^3\rangle_c\langle\delta_R\rangle_c -3\langle\delta_R^2\rangle_c^2 -6\langle\delta_R^2\rangle\langle\delta_R\rangle^2 - \langle\delta_R\rangle_c^4\\\
\dots&&
\ea
and the reduced cumulants are defined as 
\be
S_p(R)\equiv\frac{\langle\delta_R^p\rangle_c}{\langle\delta_R^2\rangle_c^{p-1}}.
\ee
One may define a generating function for the $S_p$ by
\be
\label{Ssum}
S(y)=\sum_{p=2}^{\infty}S_p(R)\frac{(-1)^{p-1}}{p!}y^p\;.
\ee
Then an exact expression for the PDF in terms of the cumulants is given by 
\be
\label{cumulantPDF}
P(\delta_R)d\delta_R=\frac{d\delta_R}{2\pi i}\frac{1}{\sigma_R^2}\int_{-i\infty}^{i\infty}dy\exp\left[\frac{y\delta_R}{\sigma_R^2}-\frac{S(y)}{\sigma_R^2}\right]
\ee
where we have defined $\sigma_R^2=\kappa_2$. Using the saddle point approximation in Eq.(\ref{cumulantPDF}) and collecting terms of the same order, one arrives at the Edgeworth expansion:
\be
\label{Edgeworth}
P(\nu)d\nu=\frac{d\nu}{\sqrt{2\pi}}e^{-\nu^2/2}\left[1+\sigma_R\frac{S_3(R)}{6}H_3(\nu)+\sigma_R^2\left(\frac{S_4(R)}{24}H_4(\nu)+\frac{S_3(R)^2}{72}H_6(\nu)\right)+\dots\right]
\ee
where $\nu=\delta_R/\sigma_R$ and the $H_n$ are Hermite polynomials
\ba
H_3(\nu)&=&\nu^3-3\nu\\\nonumber
H_4(\nu)&=&\nu^4-6\nu^2+3\\\nonumber
H_6(\nu)&=&\nu^6-15\nu^4+45\nu^2-15.
\ea
If only the first few terms in the expansion are kept the recovered probability distribution will be approximate and have a limited range of validity. For example, the PDF can develop negative regions if only the first $S_3$ term is kept. To use the Edgeworth expansion for given values of $\delta_R,\sigma_R,S_3(R) \dots $ we will therefore have to check that we remain in a region where the PDF is well behaved. This is discussed in more detail in Appendix \ref{Distributions}.

\begin{figure}[h]
\begin{center}
$\begin{array}{cc}
\includegraphics[width=0.4\textwidth,angle=0]{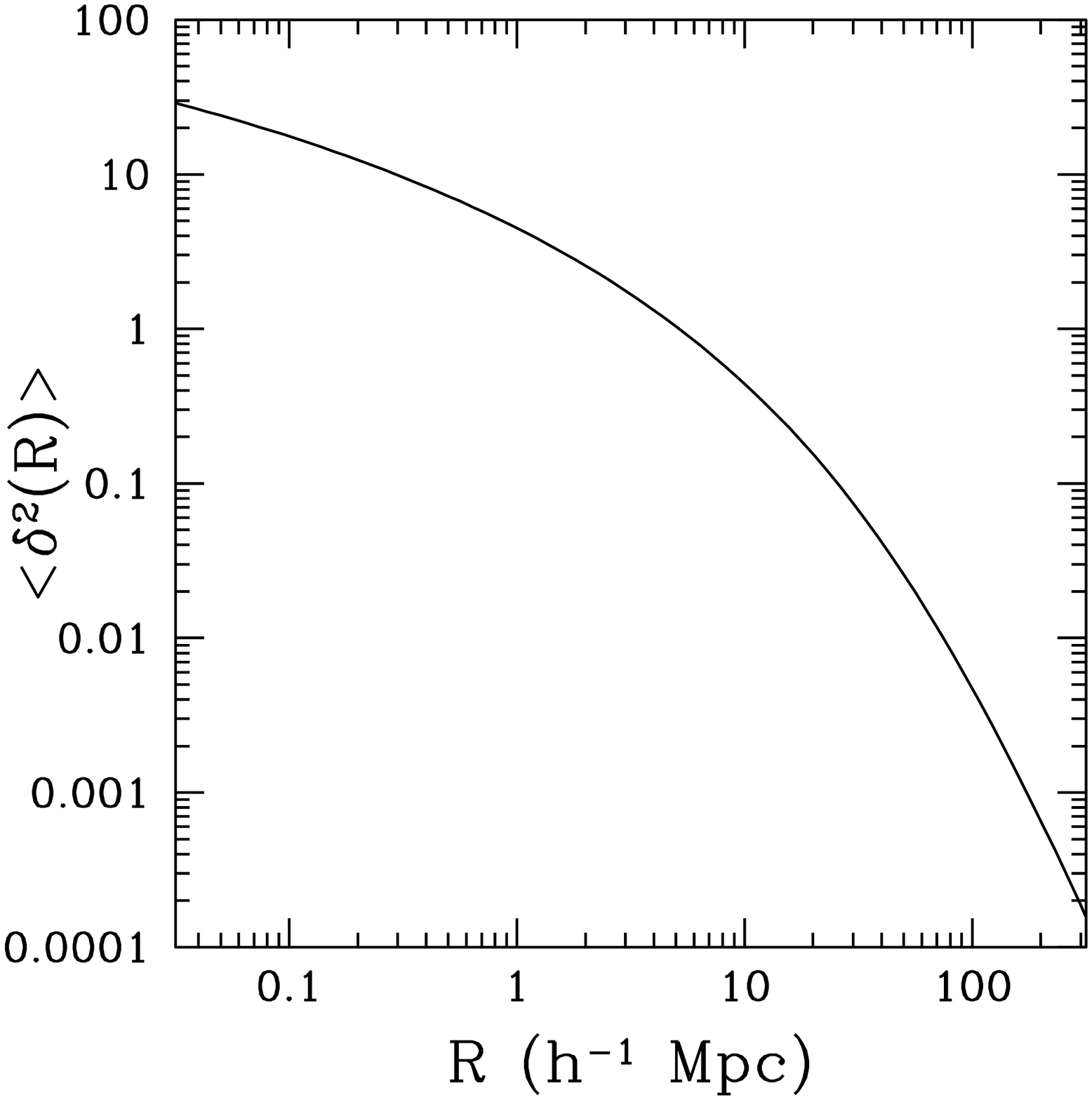} &
\includegraphics[width=0.4\textwidth,angle=0]{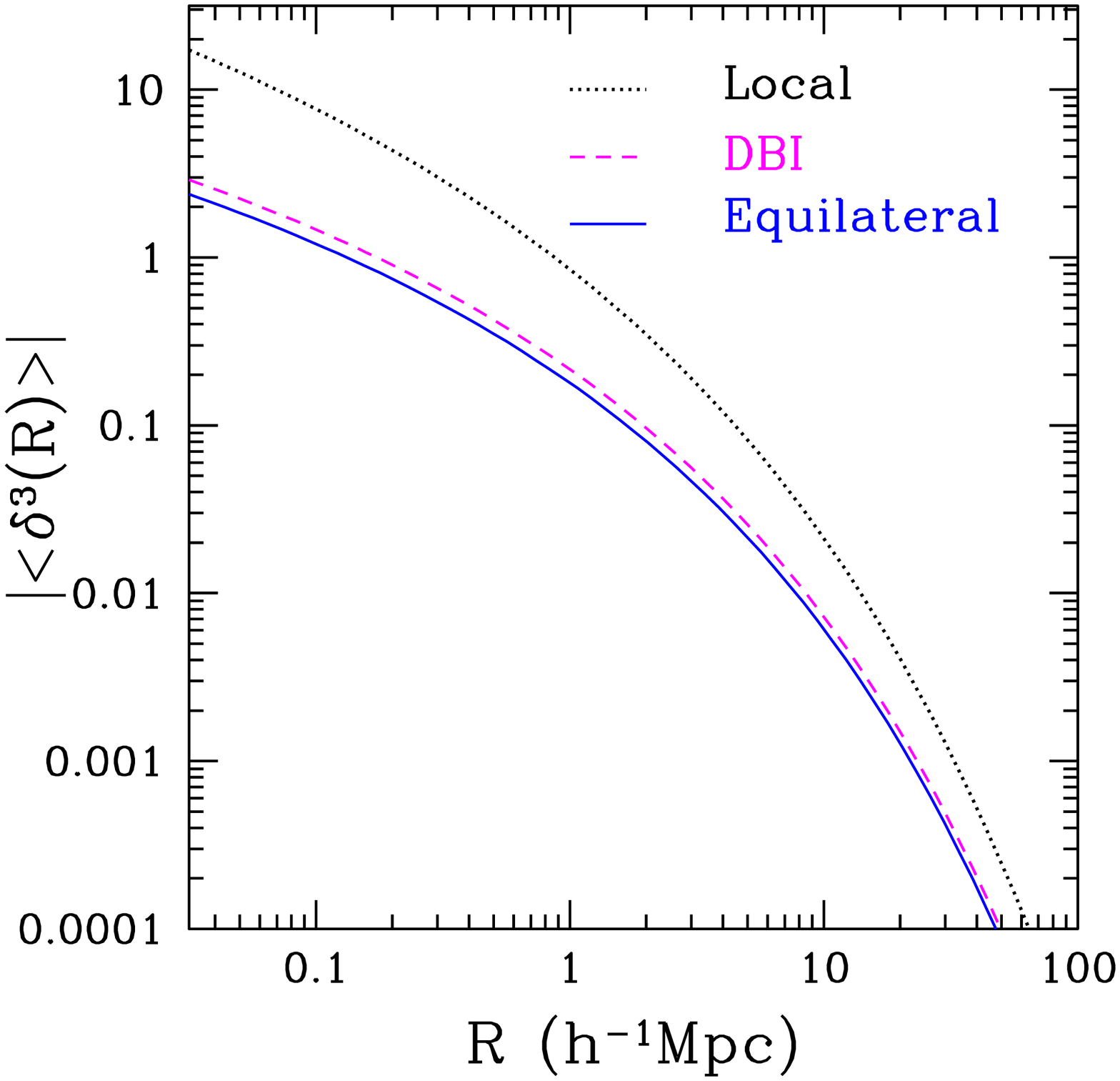} \\
\mbox{(a)} &\mbox{(b)}\\
\includegraphics[width=0.4\textwidth,angle=0]{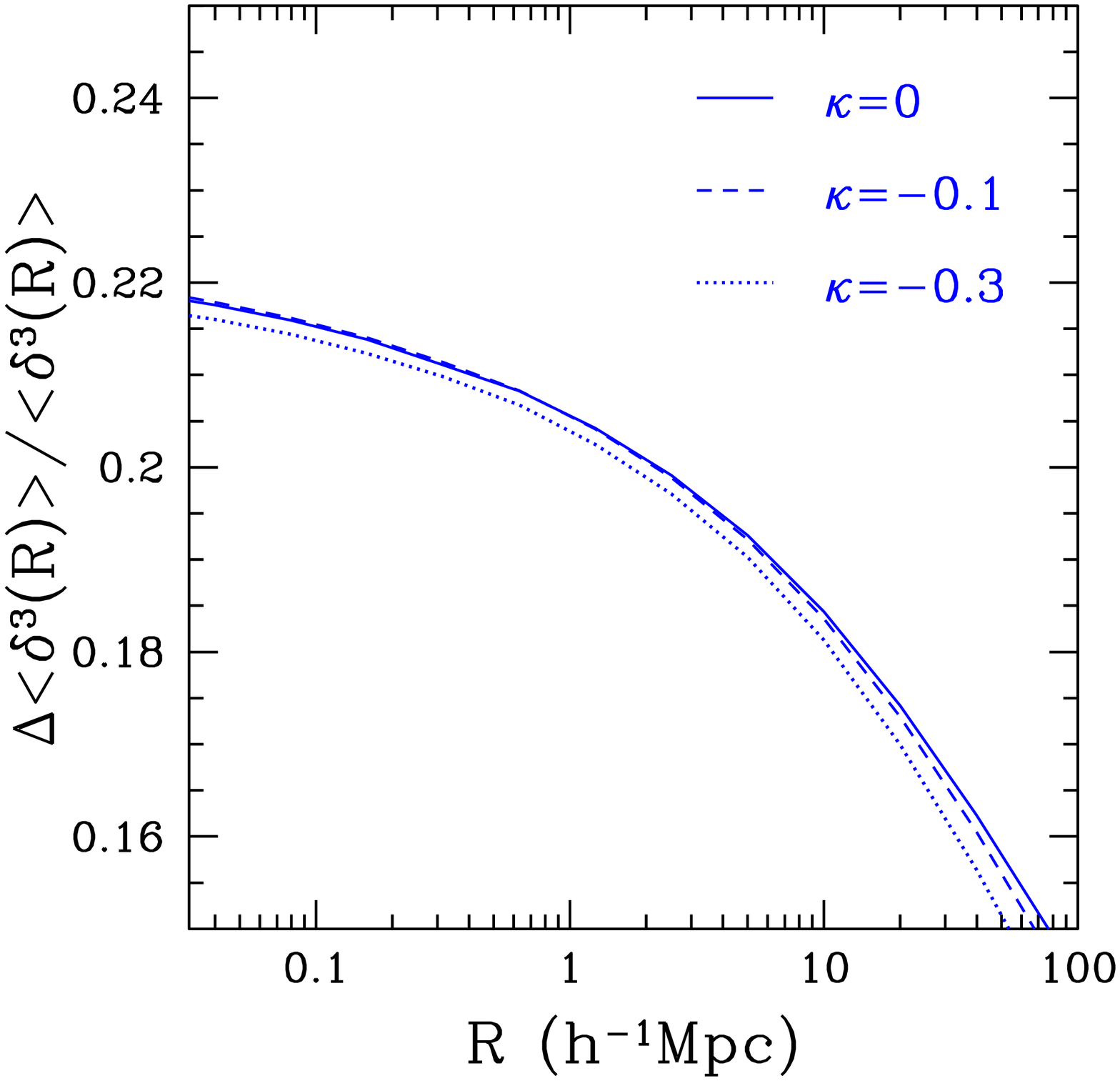} &
\includegraphics[width=0.4\textwidth,angle=0]{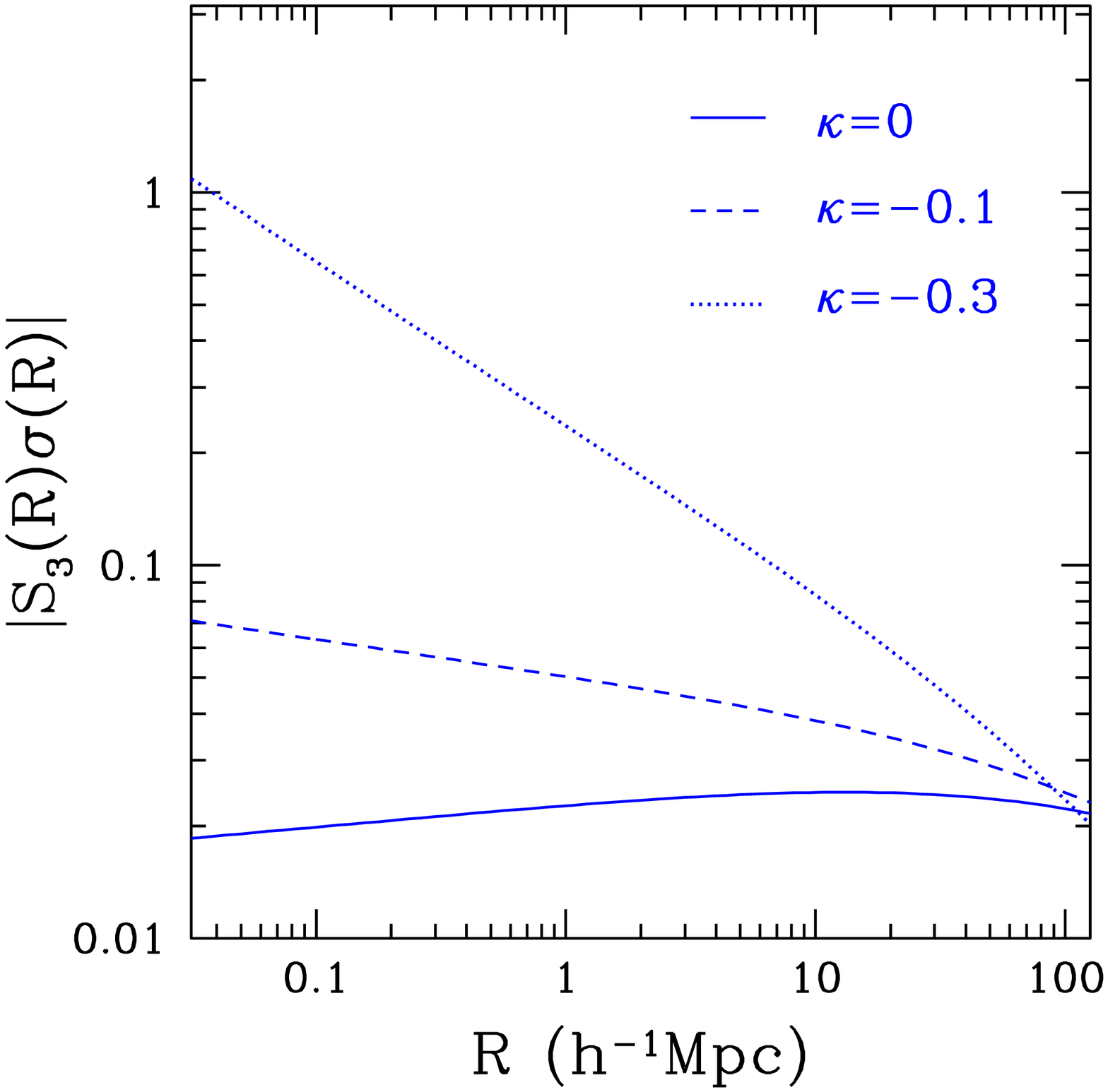} \\
\mbox{(c)}&\mbox{(d)}
\end{array}$
\caption{(a) The smoothed variance. (b) The smoothed skewness for the local, equilateral and DBI type models all with $f^{eff}_{NL}=-256$ and $\kappa=0$. (c) The fractional difference between the smoothed skewness for the DBI-type model and that for the scale-dependent equilateral model with several values of the parameter $\kappa$. (d) The scale-dependence of the non-Gaussianity is visible in $S_3\sigma=\VEV{\delta^3}/\VEV{\delta^2}^{3/2}$ for the DBI model (just the $\mathcal{A}_c$ term).}
\label{deltasfig}
\end{center}
\end{figure}

The variance of the smoothed density fluctuation is
 \ba
\sigma^2(R)=\VEV{\delta^2_R}&=&\dthreek{}\int\frac{d^3{\bf k^{\prime}}}{(2\pi)^3} W_R(k)W_R(k')M(k,z)M(k',z)\VEV{\zeta(\vk)\zeta(\vk')}\nn
&=&\int_0^\infty \frac{dk}{k} W_R(k)^2M(k,z)^2 \mathcal{P}^\zeta(k)
\ea
where we have used Eq.(\ref{Mka}) to relate $\delta$ and $\zeta$. The smoothed variance for a (real space) top hat window function is shown in Figure \ref{deltasfig}. The smoothed skewness is calculated from the three-point function
\ba
\VEV{\delta^3_R}&=&\dthreek{1}\dthreek{2}\dthreek{3}W_1W_2W_3M_1M_2M_3\VEV{\zeta(\vk_1)\zeta(\vk_2)\zeta(\vk_3)}\nn
&=&\dthreek{1}\dthreek{2}W_1W_2W_{12}M_1M_2M_{12}(2\pi)^4(\mathcal{P}^{\zeta}(K))^2\frac{\mathcal{A}(k_1,k_2,k_{12})}{k_1^3k_2^3k_{12}^3}
\label{d3Req}
\ea
where $k_{12}=\sqrt{k_1^2+k_2^2+2\vk_1\cdot\vk_2}$ and $K=k_1+k_2+k_3$. We have collapsed the $\delta$-function in Eq.(\ref{SS3pt}) setting $\vk_3=\vk_1+\vk_2$ to get the second line, though in practice it will be computationally easier to use the $\delta$-function to collapse different integrals depending on the form of the integrand (since $\mathcal{A}$ is a sum of terms). 

The smoothed skewness for the local, equilateral and $\mathcal{A}_c$ models is shown in Figure \ref{deltasfig}. As expected, the smoothed skewness for the local model is quite different in both shape and amplitude from the equilateral and $\mathcal{A}_c$ models. In Figure \ref{deltasfig} we also show the fractional error between the equilateral and $\mathcal{A}_c$ forms of the smoothed three-point function, $\left(\VEV{\delta^3_{c}(R)}-\VEV{\delta^3_{eq}(R)}\right)/\VEV{\delta^3_{eq}(R)}$. On the scales we are interested in the assumption of the equilateral form leads to an error of roughly $20\%$ for several different values of $\kappa$. Since the error is slowly varying across the range of scales clusters probe, one can imagine scaling constraints on $f_{NL}^{eq}$ to infer constraints on $f_{NL}^c$. 

We will use Eq.(\ref{Edgeworth}) together with the smoothed cumulants for the various models to predict the effect of scale-dependent non-Gaussianity on cluster counts.

\subsection{The Non-Gaussian Mass Function}
\label{NGMF}
The halo mass function  gives the number of collapsed structures (halos) at a given redshift, per unit volume with mass within a given mass interval (between M and M+dM). Expressions for the halo mass function in the presence of non-Gaussian initial conditions have been derived as extensions of the Press-Schechter formula \cite{Press:1973iz,Chiu:1997xb,Robinson:1999se, Matarrese:2000iz, Warren:2005ey, Lukic:2007fc}. Our derivation of the non-Gaussian mass function follows that of \cite{Chiu:1997xb,Robinson:1999se} but here we use the Edgeworth expansion (Eq.(\ref{Edgeworth}) keeping terms up to order $\sigma S_3$) for our non-Gaussian probability distribution. The approach is similar to that of Matarrese, Verde and Jimenez (MVJ) \cite{Matarrese:2000iz}. However, we take a different approximation in order to have a better analytic understanding of the range of validity of the expansion, this is discussed further in Appendix \ref{AppMVJ}. 

In the spirit of Press-Schechter we assert that the fraction of a volume $V$ that has collapsed into objects of mass $M$ is proportional to the probability that the density fluctuation smoothed on scale $M$ is above the threshold for collapse $\delta_c$. This implies that\footnote{ For a general probability distribution the ``fudge factor" of $2$ may need to be modified. \cite{Chiu:1997xb,Robinson:1999se} determine this factor by requiring $\bar{\rho}=\int_0^\infty M \frac{dn}{dM} dM$. For the cases we consider, this condition may be reasonably imposed as long as $\kappa\rightarrow  0$ at some small scale.}
\be
f_{c}= 2 \,P(>\delta_c,M)=2 \int_{\delta_c}^\infty d\delta P(\delta,M).
\label{fcdef}
\ee
The fraction of volume $V$ that has collapsed into objects with masses between $M$ and $M+dM$ is $|f_c(M+dM)-f_c(M)|$. This is proportional to the mass of the collapsed objects with masses between $M$ and $M+dM$ in $V$ divided by the total mass in $V$.
\be
|f_c(M+dM)-f_c(M)|=\frac{\left[n(M+dM)-n(M)\right] M V}{\bar{\rho}V}\;.
\ee
Where $dn(M)dM$ is the number density of halos with masses in the range $(M,M+dM)$. Together with Eq.(\ref{fcdef}) this gives an expression for the mass function $dn(M)/dM$ in terms of the probability distribution
\be
\frac{dn}{dM}(M,z)=-2\frac{\bar{\rho}}{M}\frac{d}{dM}\left[\int_{\delta_c/\sigma(M)}^\infty\!\!d\nu P(\nu,M)\right]\;,
\label{dndmdef}
\ee
where $\bar{\rho}=\Omega_m\rho_{crit}$ is the average (comoving) matter density. The redshift dependence is carried by the threshold for collapse $\delta_c(z) \approx 1.686 \,D(z=0)/D(z)$ where $D(z)$ is the linear growth function. 

For a Gaussian probability distribution we recover the original Press-Schechter mass function

\be
\frac{dn}{dM}(M,z)=-\sqrt{\frac{2}{\pi}}\frac{\bar{\rho}}{M^2}\frac{\delta_c(z)}{\sigma_M}\frac{d\ln \sigma_M}{d\ln M}\exp[-\delta_c^2(z)/(2\sigma^2_M)]\;.
\label{massfcn}
\ee

For the non-Gaussian case, and for the skewness not too large, one can use the Edgeworth expansion for the PDF Eq.(\ref{Edgeworth}). Performing the integral in Eq.(\ref{dndmdef}) gives\footnote{If the probability distribution $P(\nu)d\nu$ is independent of mass then this step can be skipped and the mass function is just given by $\frac{dn}{dM}=-2\bar{\rho}/M\frac{\delta_c}{\sigma_M}\frac{\textrm{dln}\sigma_M}{dM}P(\delta_c/\sigma_M)$ recovering the expression introduced in \cite{Chiu:1997xb}.}
\be
P(>\delta_c|z_c,M)=\frac{1}{2}\left[1-erf\left(\frac{\delta_c}{\sqrt{2}\sigma_M}\right)\right]-\frac{S_3(M)\sigma_M}{3!}\left(1-\left(\frac{\delta_c}{\sigma_M}\right)^2\right)\frac{e^{-\frac{\delta_c^2}{2\sigma_M^2}}}{\sqrt{2\pi}}+\dots
\ee
Then the mass function is given by
\ba
\label{massfcnNG}
\frac{dn(M, z)}{dM}&=&\\\nonumber
&&-\sqrt{\frac{2}{\pi}}\frac{\bar{\rho}}{M}e^{-\frac{\delta^2_c}{2\sigma_M^2}}\left[\frac{d\textrm{ln}\sigma_M}{dM}\left(\frac{\delta_c}{\sigma_M}+\frac{S_3\sigma_M}{6}\left(\frac{\delta_c^4}{\sigma_M^4}-2\frac{\delta_c^2}{\sigma_M^2}-1\right)\right)\right.\\\nonumber
&&\left.+\frac{1}{6}\frac{dS_3}{dM}\sigma_M\left(\frac{\delta_c^2}{\sigma_M^2}-1\right)\right]\nonumber.
\ea
For $S_3$ and all higher reduced cumulants equal to zero, this reduces to the usual expression for a Gaussian distribution. What is the effect of including skewness $S_3$ in the mass function? The exponential is unchanged with respect to the Gaussian case so we need only concern ourselves with the terms in the square brackets. For the models we consider $dS_3/dM$ has the same sign as $S_3$ on cluster scales. So if  $S_3>0$ then at the low mass end where $\delta_c/\sigma_M<1$ the number of objects is \emph{decreased}. At high masses $\delta_c/\sigma_M$ grows rapidly so $S_3 >0$ \emph{increases} the number of very massive halos. 

Comparison with numerical simulations (with Gaussian initial conditions) have shown that the Press-Schechter form of the mass function over-predicts the abundance of low mass objects and under-predicts that of high-mass objects. The discrepancy is not surprising as e.g. the spherical collapse assumption made to arrive at Eq.(\ref{massfcn}) may not hold in detail \cite{Sheth:1999su}. Sheth and Tormen \cite{Sheth:1999mn} suggested a formula that is in much better agreement with simulations, and further improvements were suggested by \cite{Jenkins:2000bv,Reed:2003sq}. Since these results have not be generalized to allow for generic non-Gaussian initial conditions, we will use a Gaussian mass function in better agreement with simulations and use the Press-Schechter-derived mass function to model departures from non-Gaussianity: 
\be
\frac{dn_{NG}}{dM}(M,z)=\frac{dn_{G}}{dM}(M,z)\frac{\frac{dn_{PS}}{dM}(S_3,M,z)}{\frac{dn_{PS}}{dM}(S_3=0,M,z)}\;.
\label{dndmNG}
\ee
In the above $dn_{G}/dM$ denotes the preferred Gaussian mass function and $dn_{PS}/dM(S_3,M,z)$ denotes the non-Gaussian extension of the Press-Schechter mass function Eq.(\ref{massfcnNG}), which reduces to the original Press-Schechter expression for $S_3(M)=0$. We discuss recent N-body simulations that have investigated the validity of this approach in the conclusions.

In our calculations we take the Sheth and Tormen mass function as $dn_{G}/dM$ in Eq.(\ref{dndmNG}). The Sheth-Tormen mass function, obtained from Gaussian initial conditions and calibrated on numerical simulations,  is given by \cite{Sheth:1999mn}
\be
\frac{dn_{ST}}{dM}=-\sqrt{\frac{2 a}{\pi}}A\left(1+\left(\frac{a\delta_c^2}{\sigma^2}\right)^{-p}\right)\frac{\bar{\rho}}{M^2}\frac{\delta_c}{\sigma}\frac{d\ln \sigma}{d\ln M}e^{-a\delta^2/(2\sigma^2)}
\label{massfcnST}
\ee
where $a=0.707$, $A=0.322184$, and $p=0.3$. 

With the mass function in hand, the number of clusters per redshift interval above some mass threshold $M_{min}$ can be calculated,
\ba
\frac{dN}{dz}(M>M_{lim})&=& f_{sky} \frac{dV(z)}{dz}\int_{M_{min}}^\infty\! dM \frac{dn}{dM}(M,z)
\ea
where $f_{sky}$ is the fraction of the sky being observed and the volume element is given by
\be
\frac{dV}{dz}=\frac{4\pi}{H(z)}\left[\int_0^z \frac{dz'}{ H(z')}\right]^2.
\ee

\subsection{Predictions for Cluster Number Counts}
Galaxy clusters are the largest virialized cosmological structures in the universe, and therefore provide a unique way to explore non-Gaussianity on scales much smaller than those accessible to CMB anisotropy measurements. In addition, probes of non-Gaussianity using cluster number counts are not subject to the same set of systematic effects as CMB measurements are, so this method provides a valuable complement to the CMB. Upcoming large mass-limited surveys of galaxy clusters with SPT \cite{Ruhl:2004kv}, ACT \cite{Kosowsky:2004sw}, and Planck \cite{Geisbuesch:2006fr} will yield data sets particularly well suited to the joint analysis of cluster number counts with CMB measurements to probe scale-dependent primordial non-Gaussianity. 

\begin{figure}[th]
\begin{center}
\includegraphics[width=0.7\textwidth,angle=0]{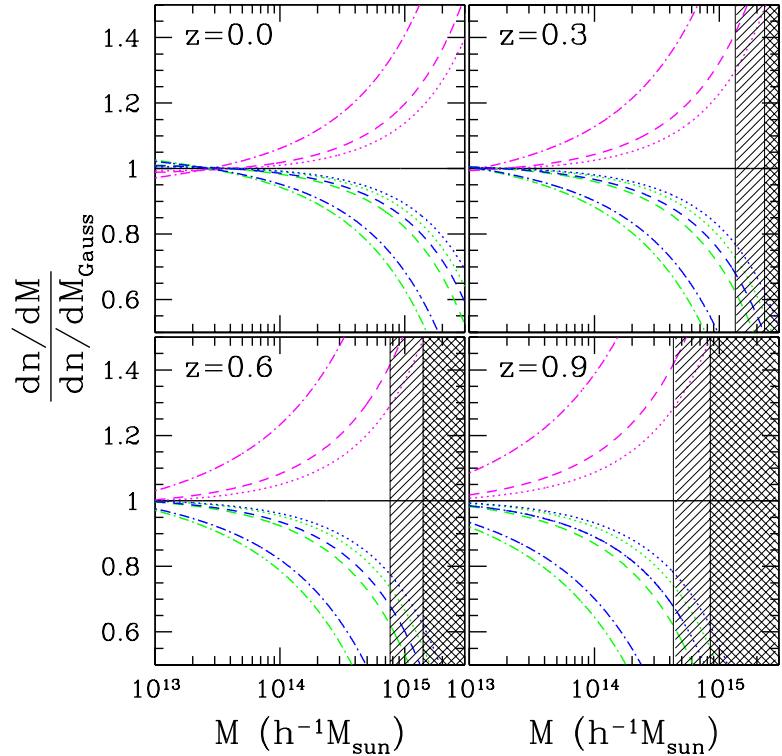} 
\caption{The ratio of the non-Gaussian mass function to the Gaussian mass function for DBI inflation (green curves, just the $c$ term) with a sound speed that saturates the bound on non-Gaussianity at the CMB scales, and for the equilateral shape of the bispectrum with $f_{NL}^{eq}(k_{CMB})=332$ (magenta upper curves) and  $f_{NL}^{eq}(k_{CMB})=-256$ (blue lower curves, showing smaller deviation from Gaussian than the $c$-term). The solid horizontal line is the Gaussian prediction, the dotted curves have no running of the non-Gaussianity ($\kappa=0$),  the dashed curves have non-Gaussianity that increases on small scales $\kappa=-0.1$ and the dot-dashed curves have $\kappa=-0.3$. The shaded regions show the regimes in which corrections to the non-Gaussian mass function from the $(S_3\sigma)^2$ term reach $5\%$ -- that is, in the shaded regions the validity of truncating the Edgeworth expansion at the first term is uncertain (see Appendix B). The validity of the expansion depends on the magnitude of the skewness: the left hand boundary is where the mass function for DBI-type with $\kappa=-0.1$ becomes invalid, the right cross-hatched region is where the mass function for $f_{NL}^{eq}(k)=332$ and $\kappa=0$ breaks down. All other curves become invalid somewhere between the two boundaries, except the $\kappa=-0.3$ cases, which becomes invalid at lower mass but where the deviation from Gaussianity is larger than shown in the range of the plot. For example, for the equilateral model with $f_{NL}=-256$, $\kappa=-0.3$, and $z=0$, the expansion is valid for $M<1.55\times10^{15}h^{-1}M_{sun}$; at $z=0.9$ the same curve is unreliable above $M=2.25\times10^{14}h^{-1}M_{sun}$.}
\label{nofMratio}
\end{center}
\end{figure}
\label{predictions}

\begin{figure}[!h]
\begin{center}
$\begin{array}{cc}
\includegraphics[width=0.5\textwidth,angle=0]{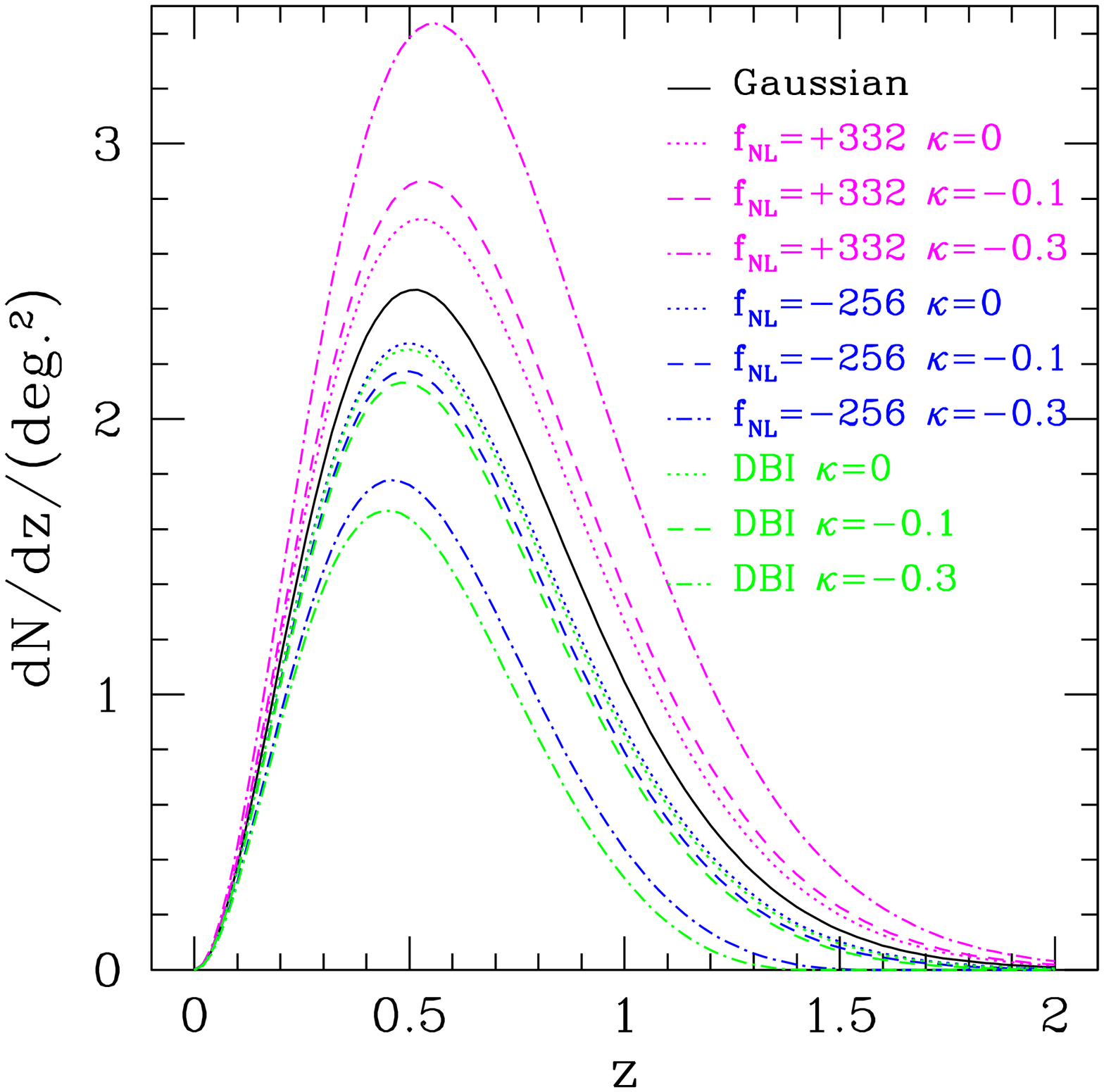} &
\includegraphics[width=0.5\textwidth,angle=0]{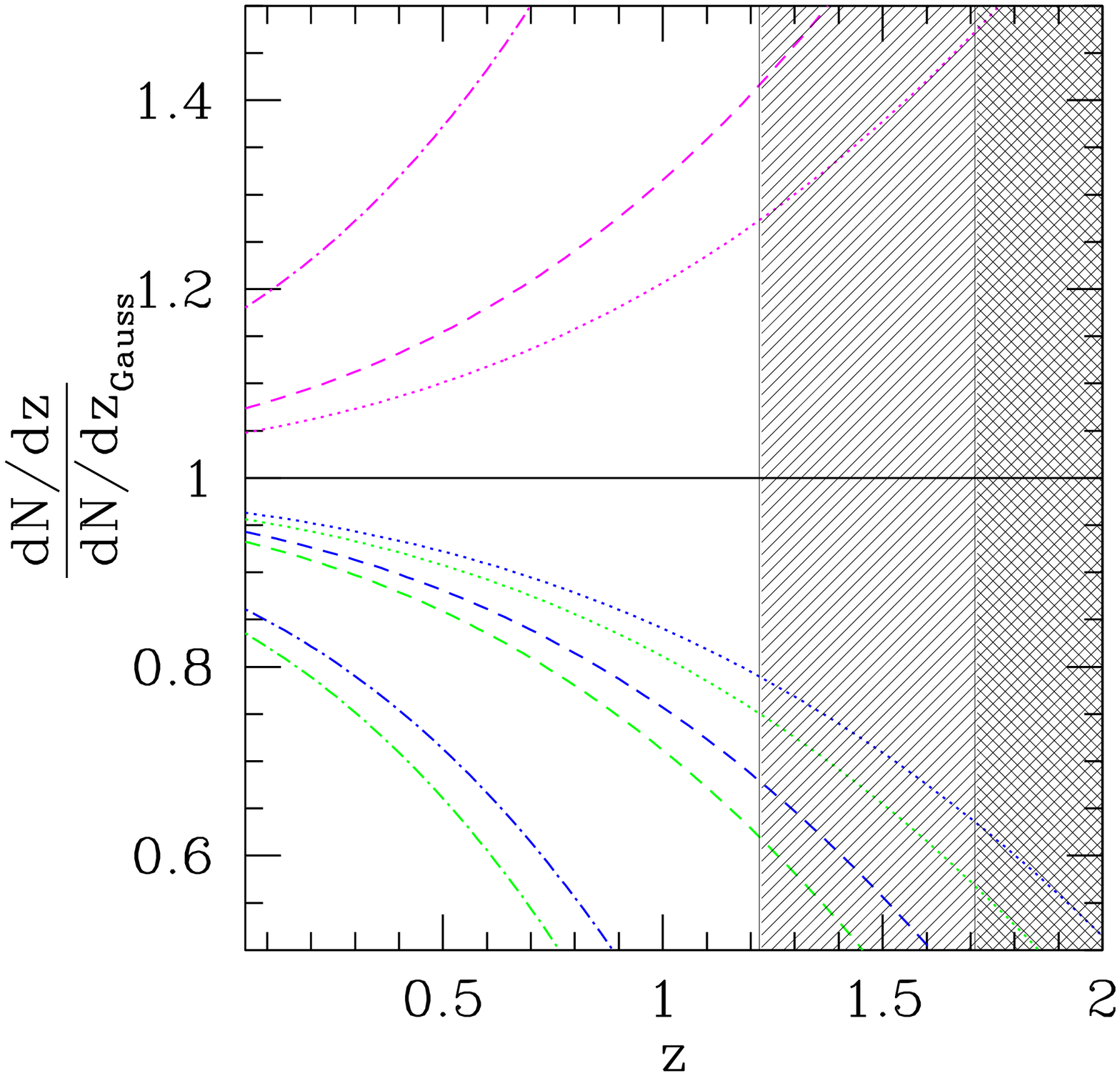} \\
\mbox{(a)} &\mbox{(b)}\\
\includegraphics[width=0.5\textwidth,angle=0]{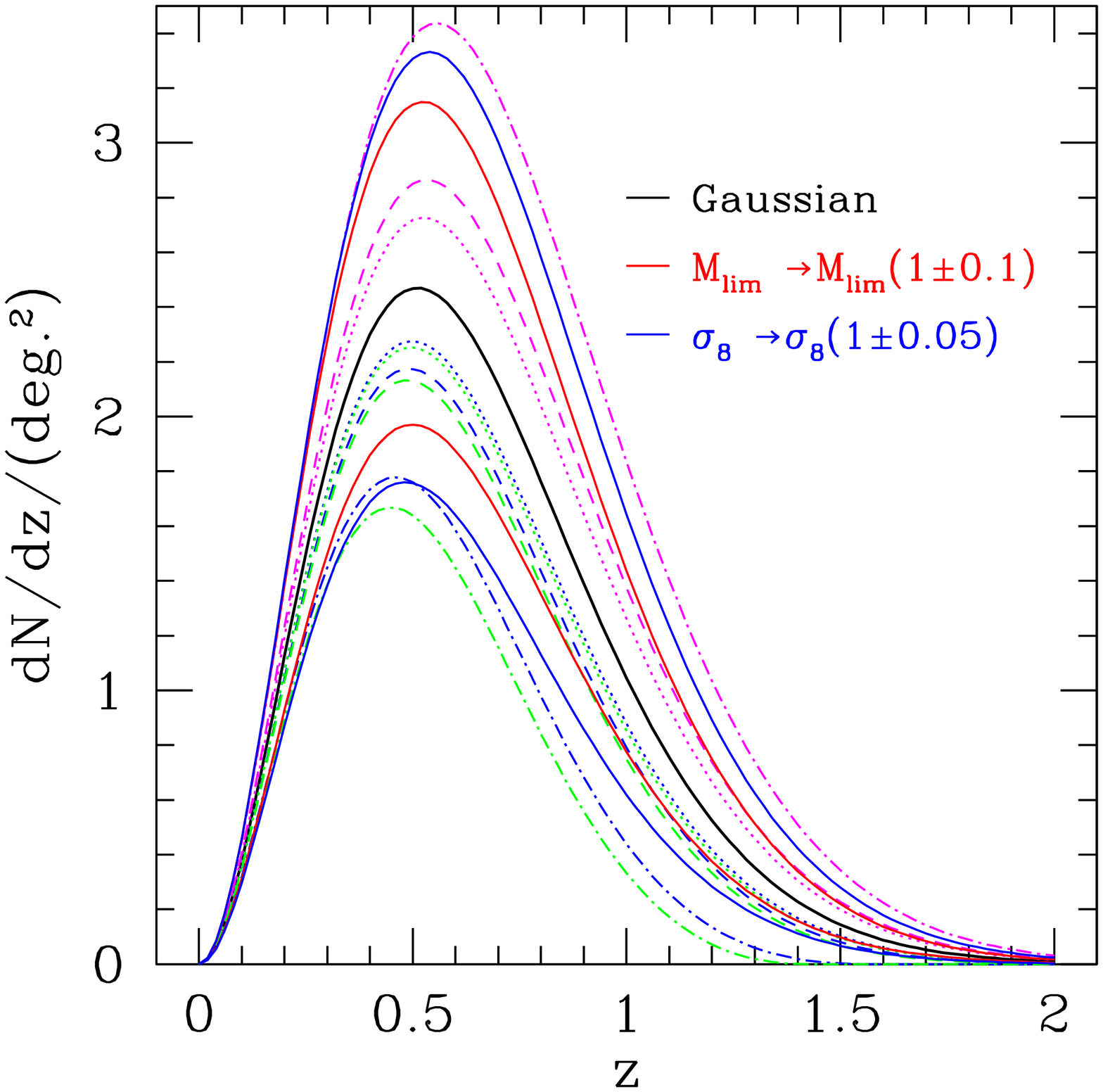} &
\includegraphics[width=0.5\textwidth,angle=0]{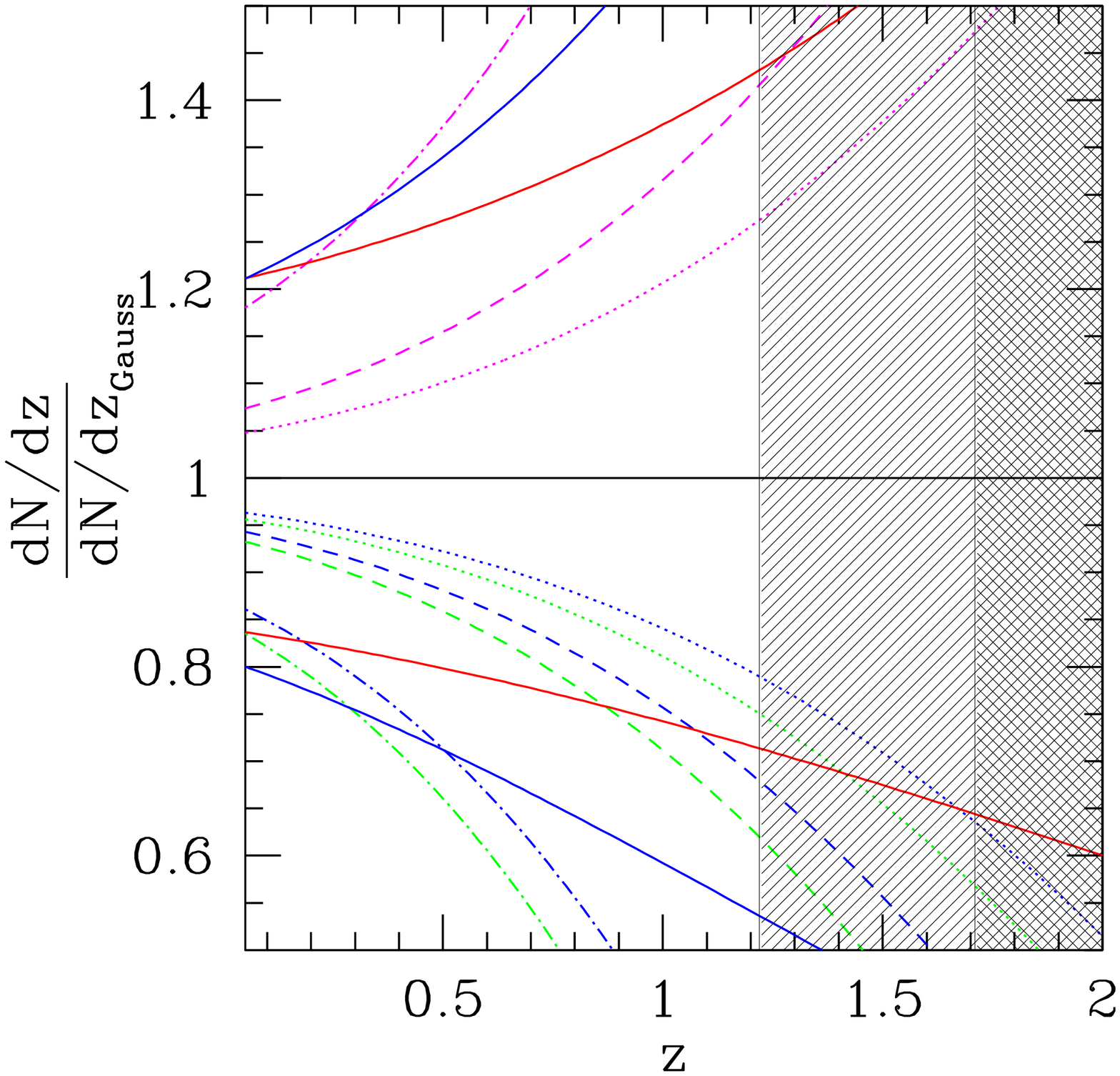} \\
\mbox{(c)} &\mbox{(d)}
\end{array}$
\caption{(a) The number of clusters with $M> M_{lim}=1.75\times 10^{14}h^{-1}M_{sun}$ per redshift interval per $deg.^2$ shown for $f_{NL}^{eq}(k_{CMB})=+332$ (magenta), $f_{NL}^{eq}(k_{CMB})=-256$ (blue) and for DBI-type inflation (green curves, just the $c$-term) with a sound speed that saturates the CMB bounds. The solid curve is the Gaussian prediction, the dotted curves have $\kappa=0$, the dashed curve has $\kappa=-0.1$ and the dot-dashed have $\kappa=-0.3$. (b) The ratio of the quantities in (a) to the Gaussian $dN/dz$. Figures (c) and (d) reproduce  (a) and (b) but also show the change in $dN/dz$ for a Gaussian cosmology if the mass threshold $M_{lim}$ or $\sigma_8$ were changed. Clearly precise knowledge of these parameters is necessary to use cluster number counts to constrain primordial non-Gaussianity. The hatched and cross-hatched regions are the same as in the previous figure.}
\label{dNdz}
\end{center}
\end{figure}
 
Given a particular form for the primordial curvature bispectrum,  Equations (\ref{d3Req}) and (\ref{massfcnNG}) can be used to calculate the non-Gaussian mass function. We consider the $\mathcal{A}_c$ term (roughly the DBI model) and the scale-dependent equilateral model (Eq.(\ref{AfNL}) and Eq.(\ref{runfNL})). For the $\mathcal{A}_c$ term we take the sound speed to be determined by Eq.(\ref{csfNLdef}),  for the equilateral model we take $f_{NL}(k_{CMB})=-256$ and $332$ (the extrema of the current $2$-$\sigma$ bounds from WMAP). For both $\mathcal{A}_c$ and the equilateral model we consider several values of $\kappa$ (see Eq.(\ref{runfNL})).  Figure \ref{nofMratio} illustrates the effect of primordial non-Gaussianity on the cluster mass function and Figure \ref{dNdz} shows the effect on cluster number counts with mass threshold of $M> M_{lim}=1.75\times 10^{14}h^{-1}M_{sun}$, roughly as predicted for ACT and SPT. As expected, the largest changes to the mass function are at high mass and high redshift.  One can also see that the equilateral model slightly under-predicts the effect of non-Gaussianity from small sound speed models (note that the origin of the equilateral model is a case that should be contained in the $\mathcal{A}_c$ term). Allowing for scale-dependent non-Gaussianity can dramatically increase the allowed effect of primordial skewness on the cluster mass function while remaining consistent with CMB observations. Figure \ref{dNdz} illustrates that even for values of $f_{NL}^{eq}(k_{CMB})$ that are within the CMB constraints the amplitude of $dN/dz$ is changed by as much as $40\%$. This suggests that upcoming mass-limited cluster survey data can indeed be used in conjunction with the CMB to constrain scale-dependent primordial non-Gaussianity. 

Cluster number counts are also quite sensitive to other cosmological parameters and systematic errors, so to understand the utility of cluster number counts as a constraint on scale-dependent non-Gaussianity we will need to examine degeneracies. In particular, the dominant effect of primordial non-Gaussianity is to change the expected number of clusters, however under or overestimating the mass threshold $M_{lim}$ or the value of $\sigma_8$ also produces significant changes to the amplitude of $dN/dz$. In the bottom panels of Figure \ref{dNdz} we show the effect of changing the value of $\sigma_8$ and $M_{lim}$ for a Gaussian model. The change in amplitude of $dN/dz$ caused by a  $5\%$ shift in $\sigma_8$ is significantly larger than the effect of primordial non-Gaussianity for all but the most strongly scale-dependent models of $f^{eff}_{NL}(k)$. On the other hand the redshift dependence of a non-Gaussian $dN/dz$ is quite different than a Gaussian $dN/dz$ with uncertain $\sigma_8$ or $M_{lim}$ so this can be used to disentangle the two effects.

\subsection{Forecasted Constraints}
We will use the Fisher matrix method to address the issue of degeneracies between cosmological parameters and primordial non-Gaussianity and forecast the sensitivities of future galaxy cluster surveys to simultaneously constrain these parameters. We account for systematic errors by allowing for an uncertainty in the mass threshold $M_{lim}$. While the Fisher matrix approach is only an approximation (it  makes the simplifying assumption of a Gaussian likelihood) it is nevertheless a useful first tool for producing forecasts and exploring parameter degeneracies.

The observation of a discrete number of clusters is  a Poisson process. The probability of observing $n_i$ clusters in the $i^{th}$ experimental bin (bins in redshift, mass or both, for example) is given by
\be
P_i=\frac{e_i^{n_i}}{n_i!}\exp(-e_i)
\ee
 where $n_i$ is the observed number of clusters in the $i^{th}$
 experimental bin and $e_i$ is the expected number in that bin.  For a set of $N_{bins}$ uncorrelated experimental bins the probability of observing $n_1 \dots n_{N_{bins}}$ clusters in each bin is then
 \ba
 P=\prod_{i=1}^{N_{bins}} P_i=\prod_{i=1}^{N_{bins}}\frac{e_i^{n_i}}{n_i!}\exp(-e_i)
 \ea
The expected number of clusters in each bin $e_i$ is a function of the model parameters (e.g. $\Omega_m$, $f_{NL}$, \dots). The maximum likelihood estimates for these parameters are those values that maximize the probability distribution $P$. We identify the likelihood with $-2$ times the logarithm of the probability distribution of the ``data", i.e. the observed number of clusters in each bin. $-2\ln P$ will behave as a $\chi^2$ distribution with the number of degrees of freedom equal to the number of parameters \cite{Cash:1979vz,Holder:2001db,Verde:2001nn} . The Fisher information matrix is then given by
\ba
F_{ab}=-\langle\frac{\partial^2\ln P}{\partial p_a \partial p_b}\rangle=\sum_{i=1}^{N_{bins}}\frac{1}{e_i}\frac{\partial e_i}{\partial p_a}\frac{\partial e_i}{\partial p_b}
\ea
 where the $p_a$ are the model parameters we use the cluster counts to constrain.  
\begin{table}[!t]
\begin{center}
\begin{tabular}{|c|rl|c|c|c|c|c|}
\hline
{\bf Info.} & \multicolumn{2}{|c|}{{\bf Fiducial Model}} &${\bf \sigma_{\Omega_m}}$ & ${\bf \sigma_h}$ & ${\bf \sigma_{\sigma_8}}$ & ${\bf \sigma_{f_{NL}}}$ & ${\bf \sigma_\kappa}$\\
\hline
{\bf WMAP} &\multicolumn{2}{|c|}{} & $0.0264$ & $0.029$ & $0.046$ & $150$ & $-$ \\
\hline
{\bf WMAP + $dN/dz$} & $f^{eq}_{NL}=38$ & $\kappa=0$ & $0.0080$ & $0.029$ & $0.026$ & $150$ &  $1.69$ \\
\hline
$''$ &  $f^{eq}_{NL}=38$ &  $\kappa=-0.3$ & $0.011$ & $0.029$ & $0.032$ & $150$ &  $1.20$ \\
\hline
$''$ & $f^{eq}_{NL}=-256$& $\kappa=0$  & $0.0076$ & $0.029$ &  $0.022$ & $150$ &  $0.17$ \\
\hline 
$''$ & $f^{eq}_{NL}=-256$ & $\kappa=-0.3$ & $0.0089$ & $0.029$ &  $0.022$ & $149$ &  $0.14$ \\
\hline
$''$ & $f^{eq}_{NL}=332$ & $\kappa=0$ & $0.010$ & $0.029$ & $0.034$ & $150$ &  $0.40$ \\
\hline
$''$ & $f^{eq}_{NL}=332$ & $\kappa=-0.3$  & $0.011$ & $0.029$ & $0.034$ & $150$ &  $0.23$ \\
\hline
{\bf Planck} & \multicolumn{2}{|c|}{} & $0.0084$ & $0.011$ & $0.015$ & $40$ & $-$\\
\hline
{\bf Planck + $dN/dz$} & $f^{eq}_{NL}=38$, & $\kappa=0.0$ & $0.0058$ & $0.011$ & $0.014$ & $40$ &  $1.00$ \\
\hline
$''$ & $f^{eq}_{NL}=38$& $\kappa=-0.3$ & $0.0070$ & $0.011$ & $0.015$ & $40$ &  $0.47$ \\
\hline
$''$ & $f^{eq}_{NL}=-256$& $\kappa=0$ & $0.0053$ & $0.011$ & $0.013$ & $40$ &  $0.09$ \\
\hline
$''$ & $f^{eq}_{NL}=-256$ & $\kappa=-0.3$ & $0.0061$ & $0.011$ & $0.013$ & $40$ &  $0.04$ \\
\hline
$''$ & $f^{eq}_{NL}=332$ & $\kappa=0$ & $0.0066$ & $0.011$ & $0.015$ & $40$ &  $0.19$ \\
\hline
$''$ & $f^{eq}_{NL}=332$ & $\kappa=-0.3$ & $0.0068$ & $0.011$ & $0.015$ & $40$ &  $0.11$ \\
\hline
\end{tabular}
\caption{The forecasted $1$-$\sigma$ errors on $\Omega_m$, $\sigma_8$, $f^{eq}_{NL}(k_{CMB})$ and $\kappa$ for three equilateral type non-Gaussian fiducial models with $\Omega_m=0.24$, $h=0.73$ and $\sigma_8=0.77$. The errors are quoted for a cluster survey with one mass bin $M>M_{lim}=1.75\times10^{14}h^{-1}M_{sun}$ and full sky coverage. To determine the errors for a survey with partial sky coverage multiply the quoted error by $1/\sqrt{f_{sky}}$. We use redshift bins of width $\Delta z=0.2$, for the models with $\kappa=0.0$ we use $7$ bins up to $\bar{z}_{max}=1.3$. For those with $\kappa=-0.3$ we use $5$ bins up to $\bar{z}_{max}=0.9$. For each value of $\kappa$, the values of $z_{max}$ are chosen to stay within the regime where the mass function is valid for all three $f^{eq}_{NL}(k_{CMB})$ values (see Appendix \protect\ref{Distributions}). Note that ACT, SPT, and Planck will provide different constraints as both the sky coverage and the depth of the survey will vary -- ACT  will yield the smallest area, deepest survey of the three, and Planck will produce a full-sky survey with a higher mass limit than the other two.
}
\label{errortable}
\end{center}
\end{table}

\begin{figure}[ht]
\begin{center}
$\begin{array}{ccc}
\includegraphics[width=0.3\textwidth,angle=0]{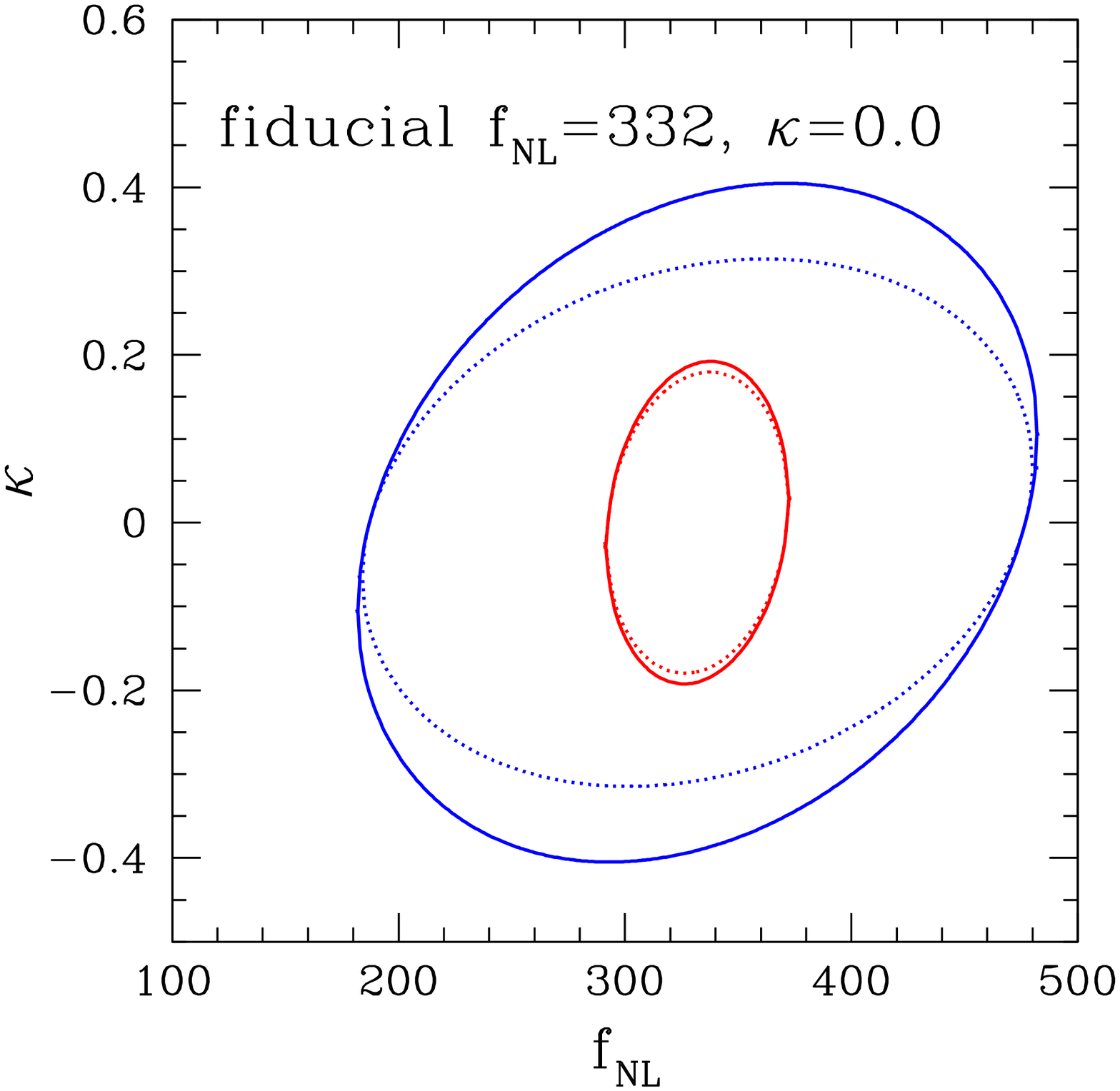} &
\includegraphics[width=0.3\textwidth,angle=0]{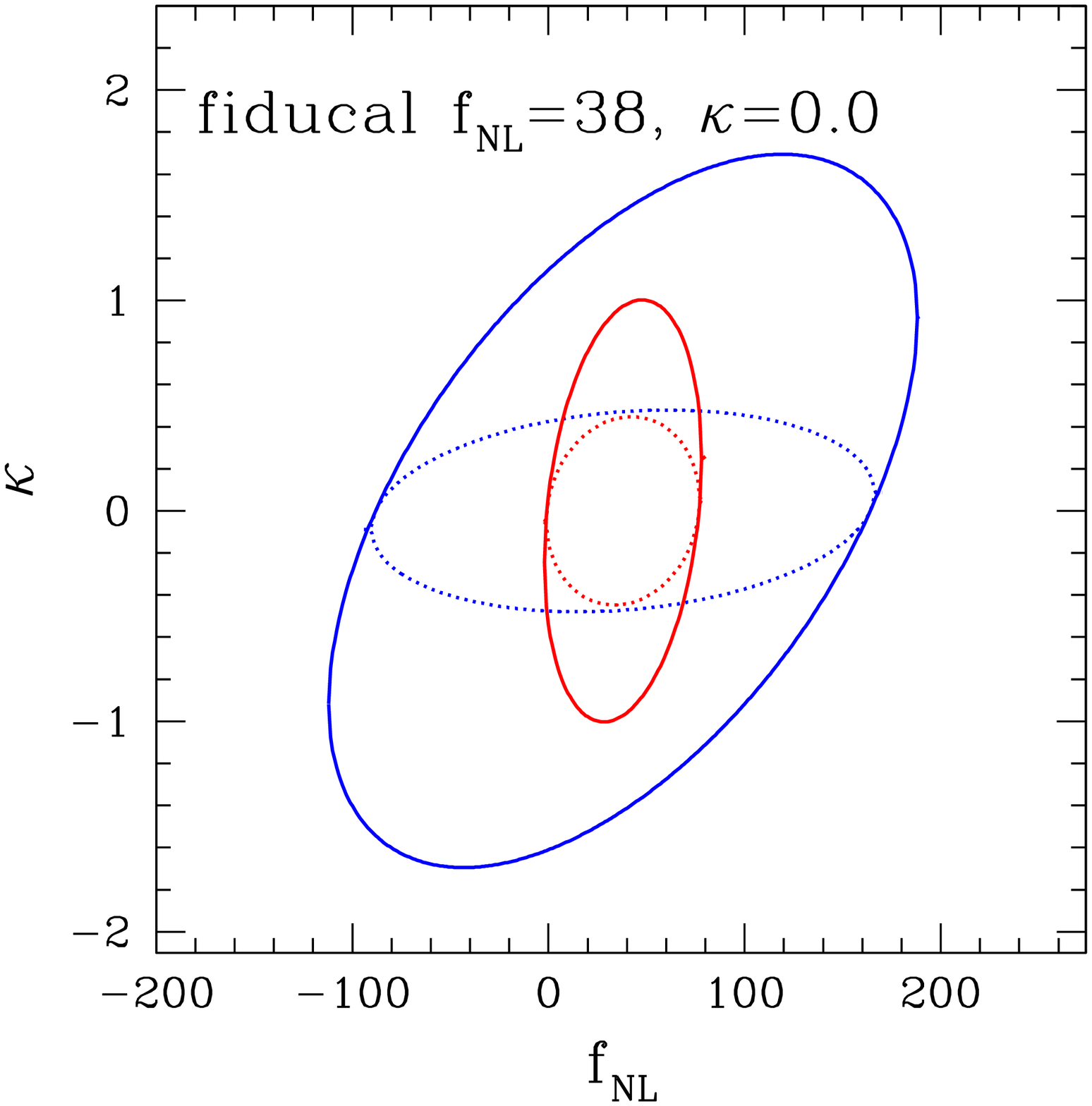} &
\includegraphics[width=0.3\textwidth,angle=0]{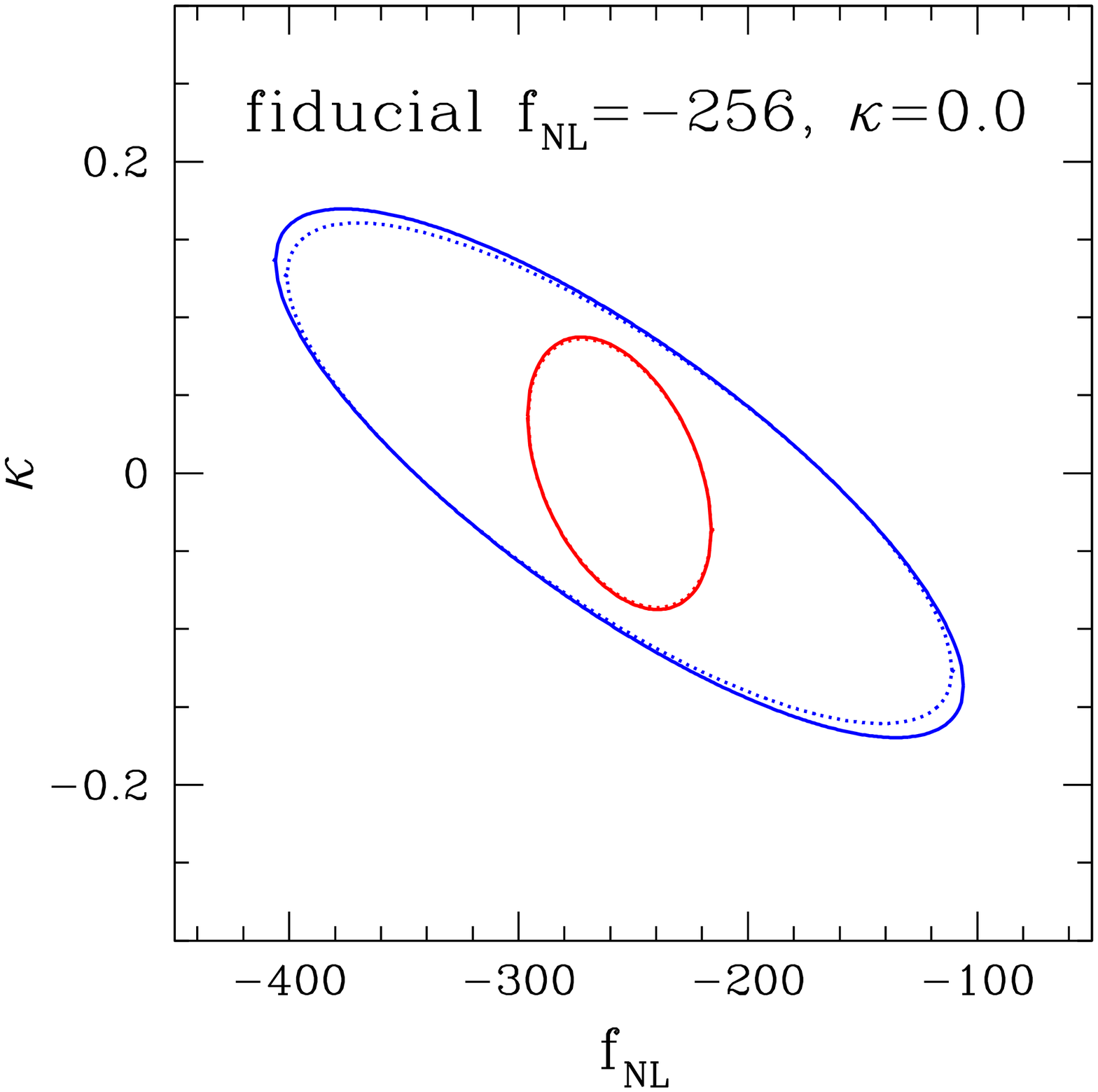} \\
\includegraphics[width=0.3\textwidth,angle=0]{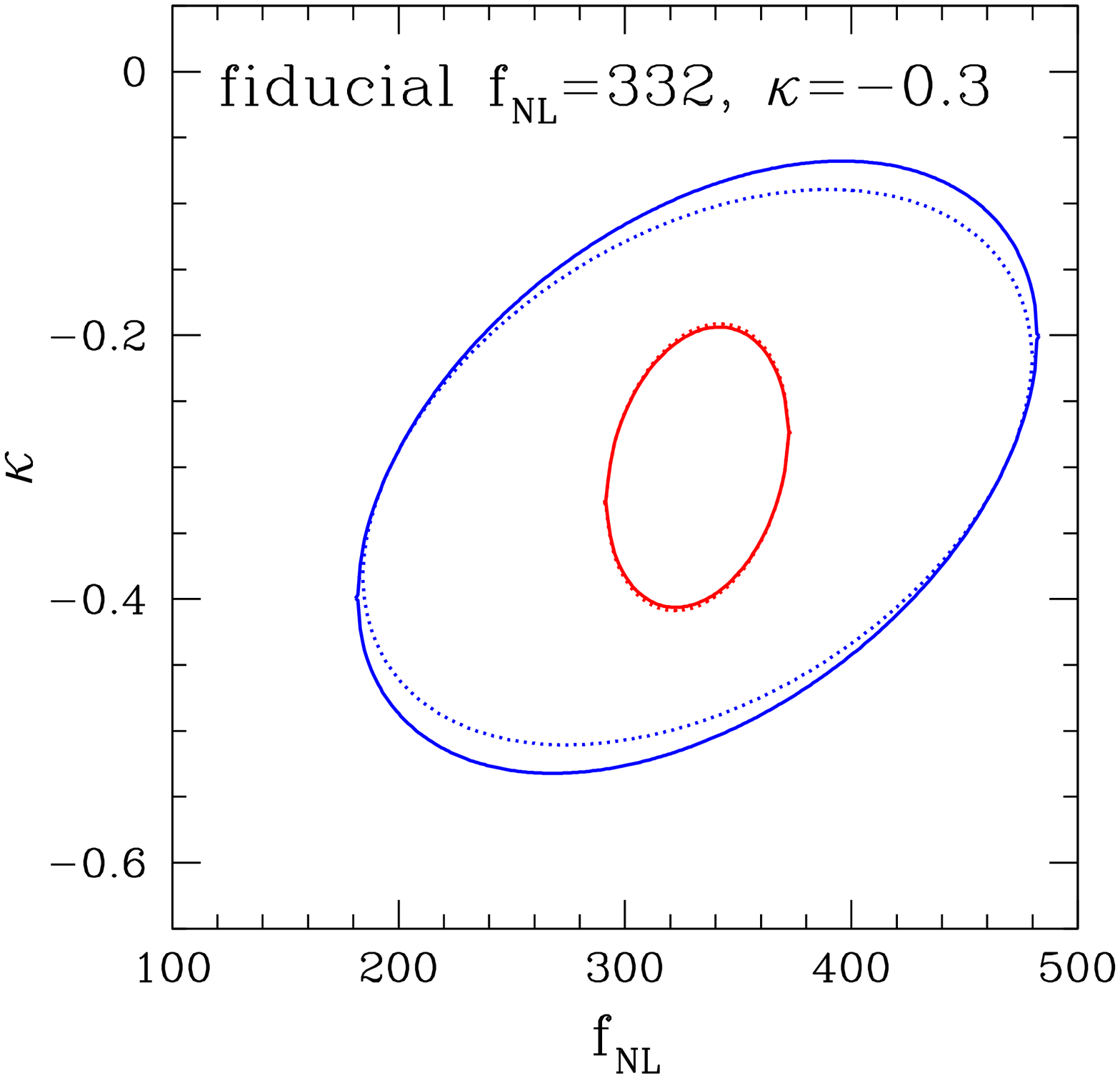} &
\includegraphics[width=0.3\textwidth,angle=0]{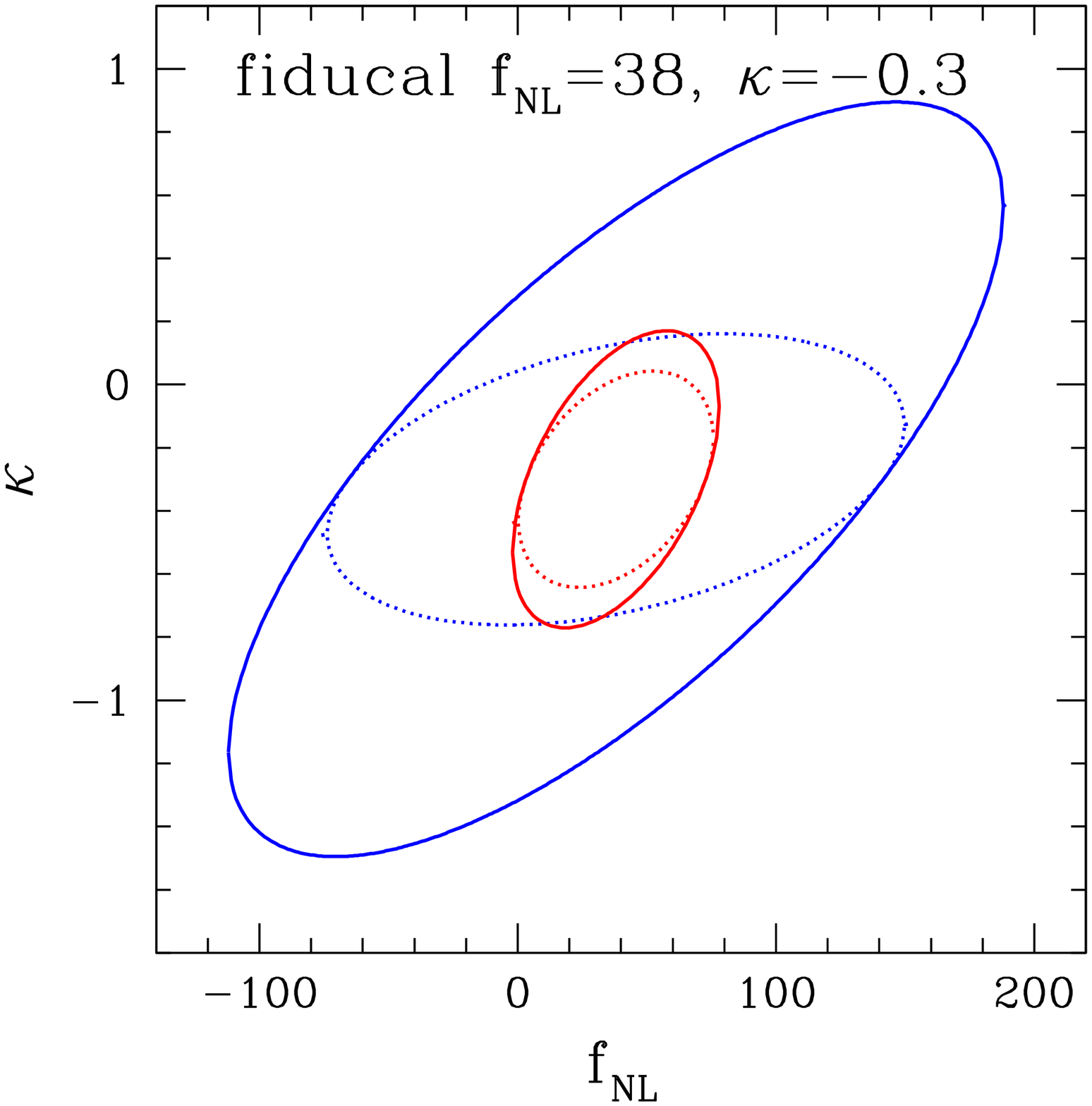} &
\includegraphics[width=0.3\textwidth,angle=0]{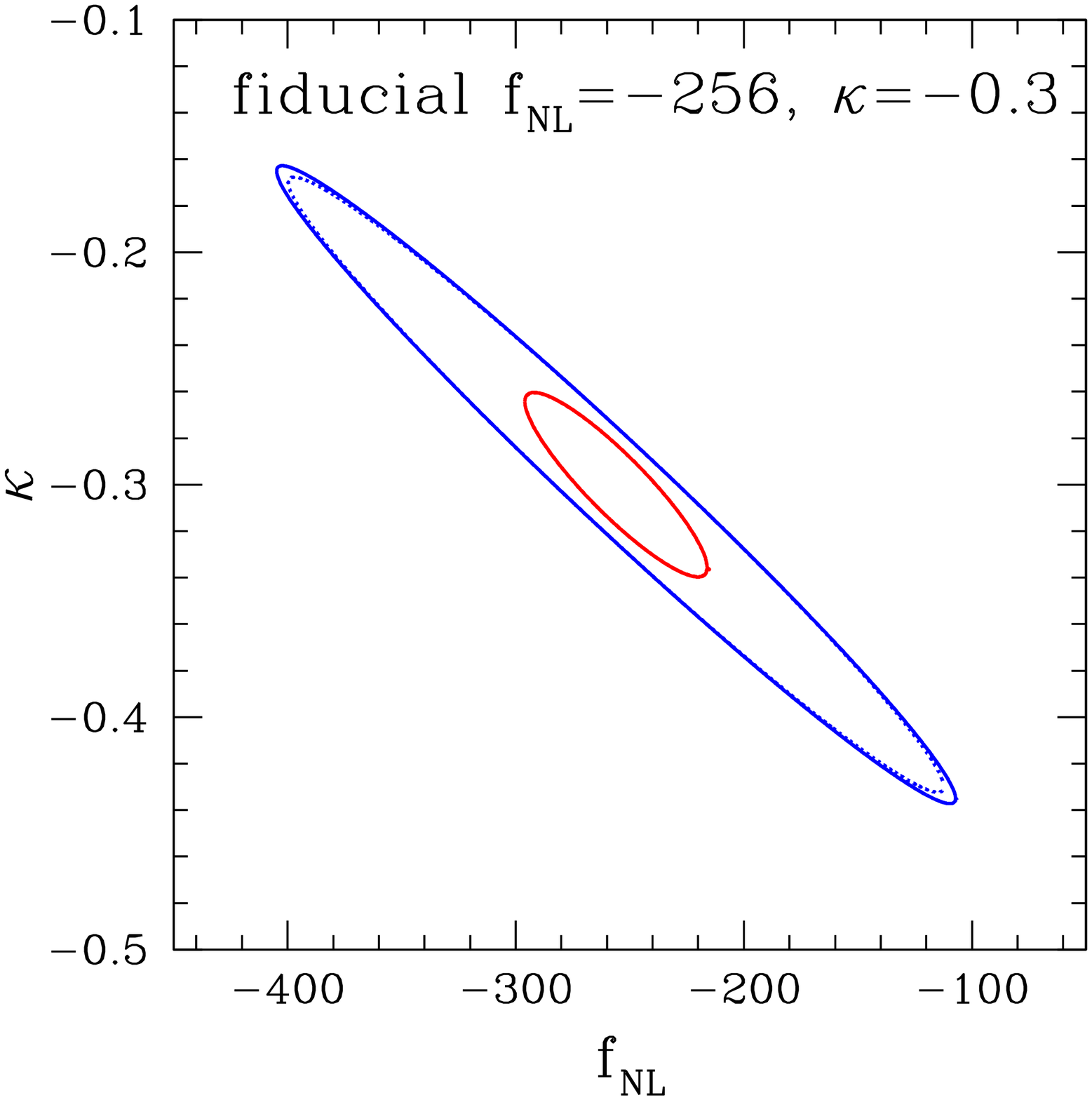} 
\end{array}$
\caption{The joint constraints on $f^{eq}_{NL}(k_{CMB})$ and the running of the non-Gaussianity $\kappa$ for a full-sky cluster survey marginalized over $\Omega_m$, $h$, $\sigma_8$ and $M_{lim}$.  Shown are $1$-$\sigma$ contours for WMAP (blue outer curves) and Planck (red inner curves) priors.  Note that the plot range varies from panel to panel. Recall, $f^{eq}_{NL}\propto k^{-2\kappa}$ so the degeneracy line changes when the sign of $f^{eq}_{NL}$ is changed. The fact that $\kappa$ is a slow roll parameter restricts its value to $\kappa << 1$, the dotted lines show the constraints if a Gaussian prior of $\pm 0.5$ is put on $\kappa$ to enforce this.}
\label{Ellipses}
\end{center}
\end{figure}

We consider the constraints on $f^{eq}_{NL}(k_{CMB})$, $\sigma_8$, $\Omega_m$, $h$, and $\kappa$. We use a single mass bin (all clusters have $M>M_{lim}=1.75\times 10^{14} h^{-1}M_{sun}$), redshift bins of width $\Delta z=0.2$ and integrate $dN/dz$ across each bin. Decreasing the bin width and increasing the total number of redshift bins does not significantly change our results. For the fiducial cosmology we assume equilateral type bispectrum with the WMAP maximum likelihood value $f^{eq}_{NL}(k_{CMB})=38$ \cite{Creminelli:2006rz}. We consider $\kappa=0.0$ and $\kappa=-0.3$ with all other cosmological parameters as given in \S \ref{clustersintro}. At high redshift corrections to our mass function become important (see Appendix \ref{Distributions}) so for the models with $\kappa=0$ we limit our analysis to bins with mean redshift $\bar{z}_i<1.3$, while for the models with $\kappa=-0.3$ we use $\bar{z}_i<0.9$. We consider two sets of priors on $\Omega_m$, $h$, $\sigma_8$ and $f^{eq}_{NL}$. The WMAP forecast assumes an $11\%$ prior on $\Omega_m$, $4\%$ on $h$, $6\%$ on $\sigma_8$ and a prior on $f^{eq}_{NL}$ of $\pm 150$ roughly reflecting the current state of knowledge from WMAP. The second forecast assumes Planck-like expectations of $3.5\%$ prior on $\Omega_m$, $1.5\%$ on $h$, $2\%$ on $\sigma_8$ and a prior on $f^{eq}_{NL}$ of $\pm 75$ \cite{Planck,Creminelli:2006rz}. A major source of uncertainty for cluster surveys is uncertainty in the mass threshold $M_{lim}$. We put a $10\%$ prior on $M_{lim}$ and all quoted constraints marginalize over $M_{lim}$. The forecasted constraints are listed in Table \ref{errortable}, for comparison we also list the constraints for fiducial models with $f_{NL}(k_{CMB})=-256$  and $f_{NL}(k_{CMB})=+332$ (the WMAP III $95\%$ confidence bounds) and all other parameters unchanged. 

A few features are apparent: while the constraints on $\Omega_m$ and $\sigma_8$ are improved beyond the prior information, the constraints on $h$ and $f_{NL}^{eq}(k_{CMB})$ are not improved by adding cluster information. What is gained by the inclusion of cluster counts is a constraint on the parameter quantifying the running of the non-Gaussianity, $\kappa$. Models with larger fiducial $f_{NL}^{eq}(k_{CMB})$ or $\kappa$ allow for better constraints on $\kappa$ while for the model with smaller $f_{NL}^{eq}(k_{CMB})$ and $\kappa=0$ clusters provide weaker constraints on $\kappa$. 

 Another point is that the constraints on the cosmological parameters $\Omega_m$, $h$, and $\sigma_8$ depend on the magnitude of $f_{NL}^{eq}$. This not too surprising because the constraints for each parameter $p_a$ depend on the magnitude of $de_i/dp_a$ and on $e_i$ for each fiducial model. The derivative of mass function, Eq. (\ref{massfcnNG}), in the presence of non-Gaussian initial conditions will have a sum of terms, some of which are dependent upon $S_3$. Since $S_3 \propto f_{NL}^{eq}(k_{CMB})$,  changing $f_{NL}^{eq}$ changes the sign and relative magnitude of these terms and therefore the magnitude of $de_i/dp_a$. On the other hand $f_{NL}^{eq}(k_{CMB})$ affects the number of clusters $e_i$ and the error on $p_a$ is inversely proportional to $\sqrt{e_i}$. 

A summary of our findings can be seen in Figure \ref{Ellipses}. Our Fisher analysis shows that if $|f^{eq}_{NL}(k_{CMB})|$ is large (just within the current WMAP bounds) cluster number counts will allow one to constrain the running of the non-Gaussianity $\kappa$. If the running of the non-Gaussianity is also large ($\kappa=-0.3$) then it is likely to be detected by a complete cluster survey. On the other hand, if $f_{NL}^{eq}(k_{CMB})\approx 38$ and $\kappa$ is large, cluster number counts may provide evidence of the running of non-Gaussianity but are unlikely to yield strong constraints. Constraints may be improved by grouping the clusters into multiple mass bins, retaining some of the information about how $dn/dM$ depends differently on each parameter. However, uncertainty in cluster masses and the rapid fall-off of the mass function at high masses cause the mass bins to be strongly correlated with one another. Finally we note that if we hadn't imposed the prior information on $f_{NL}$ from the CMB, clusters could at best achieve joint constraints of $\sigma_{fNL}\sim1000$ and $\sigma_{kappa}\sim1$ (for WMAP priors, with the same survey specifications listed above). If no running of $f_{NL}$ is allowed ($\kappa \equiv 0$) this improves to be $\sigma_{f_{NL}}\sim100-500$ depending on the sign and magnitude of $f_{NL}$. 

\section{Tests of Non-Gaussianity on Sub-CMB Scales: The Evolved Bispectrum}
\label{bispectrum}
Many planned large scale structure surveys could be well suited to test the primordial non-Gaussianity through measurements of the bispectrum. A partial list of surveys that may measure the galaxy bispectrum is SDSS \cite{GalaxySurveys1}, HETDEX \cite{GalaxySurveys2},  BOSS \cite{GalaxySurveys3}, PAU \cite{GalaxySurveys4}, ADEPT \cite{GalaxySurveys5}, LSST \cite{GalaxySurveys6}, WFMOS \cite{GalaxySurveys7} and SPACE \cite{GalaxySurveys8}. A full analysis of the potential of large-scale surveys to constrain the scale-dependence of the primordial bispectrum is beyond the scope of this paper. However, allowing scale-dependence has an interesting feature which we would like to point out. The bispectrum of the density field 
\be
\VEV{\delta(\bf k_1)\delta(\bf k_2)\delta(\bf k_3)}=(2\pi)^3\delta_D(\vk_1+\vk_2+\vk_3)B(k_1,k_2,k_3)
\ee
can be computed from perturbation theory, to leading order this is
\ba
B(k_1,k_2,k_3)=B^{I}(k_1,k_2,k_3)+B^{G}(k_1,k_2,k_3)
\label{bispec}
\ea
where $B^{G}$ is the gravitationally induced bispectrum and $B^I$, the initial bispectrum, is given by
\ba
B^I(k_1,k_2,k_3,z)&=&M(k_1,z)M(k_2,z)M(k_3,z)B^\zeta(k_1,k_2,k_3)\nn
&=&(2\pi)^4M(k_1,z)M(k_2,z)M(k_3,z)\frac{\mathcal{P}^\zeta(K)^2}{k_1^3k_2^3k_3^3}\mathcal{A}(k_1,k_2,k_3)\, .
\ea
For a review of the calculation leading to Eq.(\ref{bispec}) and expressions for $B^G(k_1,k_2,k_3)$ see \cite{Bernardeau:2001qr}. 
 
\begin{figure}[h]
\begin{center}
$\begin{array}{cc}
\includegraphics[width=0.5\textwidth,angle=0]{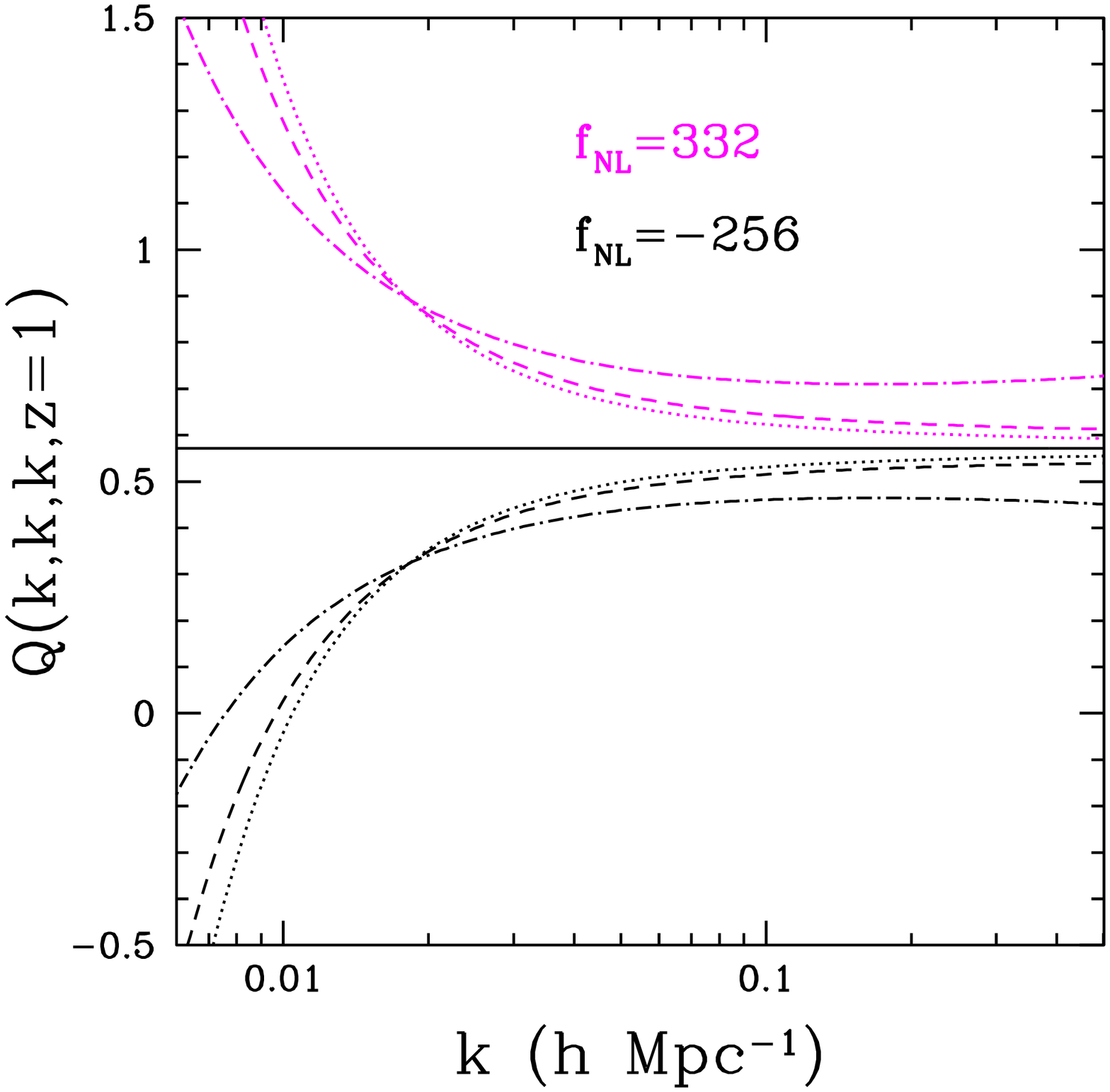} &
\includegraphics[width=0.5\textwidth,angle=0]{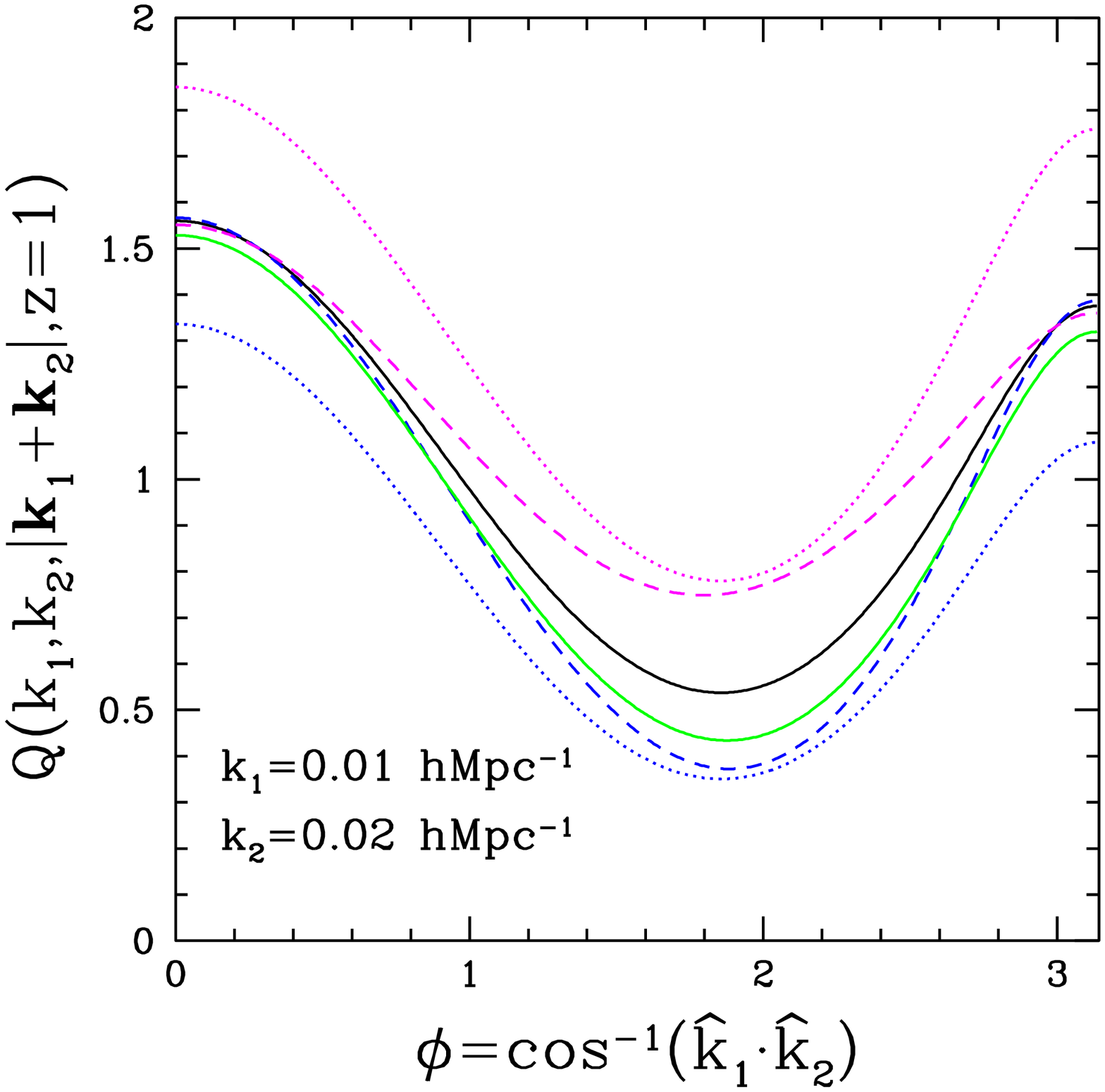} \\
\mbox{(a)} &\mbox{(b)}
\end{array}$
\caption{(a) The reduced bispectrum including both the gravitationally induced non-Gaussianity and the evolved primordial non-Gaussianity from a scale-dependent model with $f_{NL}=0$ (solid black line), $f_{NL}(k)=332(k/k_{CMB})^{-2\kappa}$ (magenta upper curves) and $f_{NL}(k)=-256(k/k_{CMB})^{-2\kappa}$ (black lower curves) shown here at $z=1$. The dotted line has $\kappa=0$, the dashed $\kappa=-0.1$ and the dot-dashed has $\kappa=-0.3$. The behavior of the dominant term in DBI with a sound speed saturating CMB bounds is identical to the $f_{NL}(k)=-256(k/k_{CMB})^{-2\kappa}$. In the equilateral limit shown above the local and equilateral models agree. (b) The reduced bispectrum as a function of $\phi$ where $\hat{k}_1\cdot \hat{k}_2=\cos\phi$. The solid black curve has no primordial contribution. Upper magenta lines have (constant) $f_{NL}=332$, dashed is equilateral shape, dotted is the local shape. The green solid curve is the dominant contribution for DBI (just the $c$-term) with $\kappa=0$. Lower blue lines have (constant) $f_{NL}=-256$, dashed is equilateral shape, dotted is the local shape.}
\label{Qtheta}
\end{center}
\end{figure}

A useful quantity is the reduced bispectrum given by 
\ba
Q(k_1,k_2,k_3,z)&=&\frac{B(k_1,k_2,k_3,z)}{P_L(k_1,z)P_L(k_2,z)+P_L(k_1,z)P_L(k_3,z)+P_L(k_2,z)P_L(k_3,z)}\\
Q(k,k,k,z)&=&\frac{4}{7}+\frac{6 \,f^{eff}_{NL}(k)}{5M(k,z)}\\
&=&\frac{4}{7}+\frac{6\, f^{eff}_{NL,CMB}(k/k_{CMB})^{-2 \kappa}}{5M(k,z)}
\label{Q}
\ea
where $P_L(k,z)$ is the linear power spectrum given in Eq.(\ref{pdef}). The first term in Eq. (\ref{Q}),  is the gravitational contribution,\footnote{For a universe with $\Omega_m\neq 1$, there are corrections of order $\Omega_m^{-2/63}-1$, we restrict our discussion to high redshift ($z=1$) where these corrections are $\sim 1\%$. The exact form is calculable, see for instance \cite{Bernardeau:2001qr}.} the second is from the primordial bispectrum. For small $k$ (small compared with the horizon scale at matter radiation equality $k_{eq}$) $M\sim k^2$, while for $k>>k_{eq}$, $M \sim \ln(k)$. Thus for a constant $f^{eff}_{NL}$, the $k$ dependence of $Q^I(k)$ is fixed, in particular $|Q^I(k)|$ is a decreasing function. This is in contrast to the scale-dependent case where the dominant term has an additional dependence on $f^{eff}_{NL}(k)$. Recall for a sound speed model the dominant terms have $f^{eff}_{NL}(k)\sim (1/c_s^2(k)-1)$ where $c_s(k)\sim k^{\kappa}$.  If, for example, $|f^{eff}_{NL}(k)|$ is an increasing function of $k$, $|Q^I_c(k)|$ can be an increasing function of $k$. This point is illustrated in the left panel of Figure \ref{Qtheta}. For models with $\kappa < 0$ the primordial contribution to the bispectrum turns over and begins to increase at some $k$.  Of course, the turnover scale may be at such large $k$ that corrections due to non-linear evolution become important and this calculation will need to be modified.

In contrast to cluster number counts, the bispectrum retains the shape-dependence of the primordial bispectrum, potentially allowing one to discriminate between the local and equilateral shapes (e.g. panel (b) of Figure \ref{Qtheta} and also \cite{Sefusatti:2007ih}). Additionally, the reduced bispectrum is independent of $\sigma_8$, which for cluster number counts is degenerate with $f_{NL}$. On the other hand, galaxy surveys measure the perturbation to the galaxy number density $\delta_{gal}$ rather than the matter density perturbation $\delta$. One often assumes that the two are related by 
\be
\delta_{gal}({\bf x})=b_1\delta({\bf x})+b_2\delta^2({\bf x}).
\ee
In \cite{Sefusatti:2007ih} it was shown that for the equilateral shape bispectrum, there is a degeneracy between $f_{NL}^{eq}$ and the bias parameters $b_1$ and $b_2$. Cluster number counts have no dependence on these parameters, so they may be useful in breaking this degeneracy. (There is no serious degeneracy between $f_{NL}$ and the dark energy equation of state, according to \cite{Sefusatti:2007ih}.) Perhaps the most serious drawback of cluster number counts, that they are sensitive only to the amplitude of $S_3$ and not to the shape of the bispectrum, can be turned into an advantage. That is, cluster number counts will be sensitive to any shape (and scaling) of non-Gaussianity that leads to a non-zero $S_3$ at cluster scales, knowledge of the particular shape of the bispectrum is less important. Finally, we mention that the systematics involved in counting collapsed objects are certainly different than those arising in measurements of correlation functions.  A multi-pronged approach should be taken.

\section{Conclusions}
\label{Conclusions}
Primordial non-Gaussianity is an excellent discriminator of inflationary models. In particular, detection of large primordial non-Gaussianity ($|f^{eff}_{NL}|$ greater than a few) would be a signature of inflationary physics outside of the smooth, slow-roll paradigm \cite{Acquaviva:2002ud, Maldacena:2002vr}. Many models that give equally viable predictions for the power spectrum can be distinguished by their non-Gaussianity, which may be an important tool for understanding the fundamental origin of inflation. Currently there is no reason to expect scale-invariant non-Gaussianity (even the slow-roll contribution is scale-dependent, although quite small in magnitude), and a number of well-motivated inflationary models predict a non-Gaussianity that can be large and run with scale \cite{Seery:2005wm, Chen:2006nt}. The possibility of scale-dependent non-Gaussianity gives new importance to observational probes at multiple scales. From a very optimistic point of view, observably large non-Gaussianity might be a signature of stringy physics, and in that case its running may probe the geometry of the extra dimensions.

In this paper, we have motivated scale-dependent non-Gaussianity which arises in string-inspired inflationary models such as DBI inflation (\S \ref{DBIexample}). In the models we discuss there are also well-defined relations between the ``running" of the non-Gaussianity and other observables such as the tensor to scalar ratio (\S \ref{scaledepsec}).  We proposed a simple power law description of scale-dependent non-Gaussianity in Eq.(\ref{runfNL}) valid for general non-Gaussian shapes. 
 
While the CMB constrains deviations from Gaussianity at the largest observable scales, the abundance of collapsed objects is sensitive to the non-Gaussianity of the primordial density field at small scales.  Allowing for scale-dependent non-Gaussianity means that models could be consistent with CMB constraints but still leave an observable imprint on other scales. For this reason we have used a new expression for the non-Gaussian halo mass function:  Eq.(\ref{massfcnNG}) including terms proportional to the skewness, and Eq.(\ref{massfcnNG2order}) expanded to higher-order cumulants. The mass function Eq.(\ref{massfcnNG}) compares well with other proposed non-Gaussian mass functions and offers an analytic understanding of the range of validity of the approximations made in the derivation of the expression (see Appendix \ref{AppMVJ}). 

We have examined the utility of cluster number counts (relevant at a scale an order of magnitude smaller than the current smallest CMB scale) to constrain the magnitude, sign, and scale-dependence of primordial non-Gaussianity. A modest running of the skewness towards small scales leads to an $f^{eff}_{NL}$ at cluster scales that is a few times larger in magnitude than $f_{NL}^{eff}$ at CMB scales. Thus, despite the fact that clusters typically provide weaker constraints on (constant) $f_{NL}$ when compared with the CMB, they can nevertheless provide important bounds on scale-dependent non-Gaussianity. For example, non-Gaussianity that is within the CMB bounds can easily cause a $10-40\%$ change in the number of clusters (Figure \ref{dNdz}). However, with upcoming experiments tight constraints on the parameter $\kappa$ (quantifying the running of the non-Gaussianity) are likely to be found only if the amplitude of the primordial contribution is just within the current CMB constraints, and/or the running of the non-Gaussianity is large (Table \ref{errortable} and Figure \ref{Ellipses}). 

The galaxy bispectrum can be used to constrain non-Gaussianity on scales between those probed by the CMB and cluster number counts. Measurements of the bispectrum are sensitive to the magnitude of the non-Gaussiainity and the running parameter, and also distinguish between the various shapes (local and equilateral, for example). Figure \ref{Qtheta} illustrates these effects.

The three types of measurements we have made use of here (cluster number counts, the galaxy bispectrum, and the CMB) are all subject to different systematics and different degeneracies. For example, while uncertainties in the mass-observable relation or the value of $\sigma_8$ are particularly troublesome for cluster surveys, the reduced galaxy bispectrum is completely independent of these parameters. On the other hand, a potentially problematic degeneracy exists between galaxy bias parameters and $f_{NL}^{eff}$, but cluster number counts are not dependent on galaxy bias. Detection of primordial non-Gaussianity will likely require confirmation from multiple probes. Ideally, measurements across a range of scales should be combined. In addition, the current WMAP data, which by itself probes a large range of scales, should be reanalyzed to look for non-Gaussianity of a more general, scale-dependent type. It is likely that the bound we have imposed on the magnitude of $f^{eff}_{NL}$ at the smallest CMB scale may be relaxed. 
 
Many of the issues in developing the probability distribution for the initial fluctuations, which limit the utility of our analysis, would be helped by further N-body simulations for more general non-Gaussian initial conditions. Three recent papers \cite{Kang:2007gs, Grossi:2007ry, Dalal:2007cu} have examined the validity of the extended Press-Schechter approach (which we have adopted here) for non-Gaussianity of the local type. Their conclusions differ, and clearly more work should be done to determine the accuracy of the mass function we have used for models that are not at all like the local model. A related issue is that we have included only the skewness in our calculations. Including higher-order cumulants may allow one to extend the regime of validity of the mass function to higher masses and redshifts.

Finally, we mention that although we have focused on the ability of near-future observations to probe scale-dependence in the primordial non-Gaussianity, one might eventually hope to do much better. For example, 21 cm observations might give an enormous number of data points at higher redshift \cite{Cooray:2006km,Pillepich:2006fj}. The smallest scale constraint may come from primordial black holes, formed by large fluctuations that re-entered very early in the radiation dominated era. Originally used as a constraint on the spectral index \cite{Carr:1975qj, Green:1997sz}, the requirement that primordial black holes that survive to the present day do not overclose the universe may also constrain the running of the non-Gaussianity. The relevant scale is roughly $k\approx10^{15}Mpc^{-1}$, many orders of magnitude beyond the range where the constraints we have investigated here apply. However, this is also far beyond the range of validity of the Edgeworth expansion ($\nu$ is large) and the contributions of higher-order cumulants cannot necessarily be neglected. We leave a full discussion of the conditions under which primordial black holes constrain the running of the non-Gaussianity for future work \cite{PBH}.

{\bf Acknowledgments} We thank Greg Bryan, Lam Hui, Jan Kratochvil, Louis Leblond, Sabino Matarrese and Sheng Wang for useful discussions. We are especially grateful to the organizers and participants of the ``Life Beyond the Gaussian" workshop where we presented an early version of this work.  A. M. gratefully acknowledges the generous support of the Alfred P. Sloan Foundation. The work of S. S. and M. L. is supported by the DOE under DE-FG02-92ER40699. M. L. is also supported by the Initiatives in Science and Engineering Program at Columbia University. L. V. is supported by FP7-PEOPLE-2007-4-3-IRG n 202182.

\appendix
\section{Review of First Order Perturbations}
\label{PerturbSign}
In this appendix we review the calculation relating perturbations in the inflaton $\delta\phi$ to primordial curvature perturbations using gauge invariant variables, and then relate the primordial curvature perturbation to perturbations in the energy density. We begin by reviewing some of the gauge invariant variables discussed in Bardeen (1980) (for a more pedagogical introduction see for instance, \cite{Mukhanov:2005sc}). We will consider scalar perturbations to the metric only, and neglect higher-order terms that appear in gauge transformations and when relating quantities using Einstein's equation. 

Begin with the following parameterization for scalar perturbations to the flat Friedmann-Robertson-Walker metric
\be
g_{00}=-(1+2A) \quad g_{0i}=g_{i0}=-a\partial_i B \quad g_{ij}=a^2(t)\left[(1+2\psi)\delta_{ij}-2\partial_i\partial_jE\right]
\ee
Bardeen's gauge invariant scalars are then
\ba
\Phi_A&=&A-\frac{\partial}{\partial t}\left[a(B-a\dot{E})\right]\nn
\Phi_H&=&\psi-aH\left[B-a\dot{E}\right]
\label{phiAH}
\ea
where $\cdot$ is the derivative with respect to $t$ and $H=\dot{a}/a$. We can also write down the gauge invariant velocity perturbation
\be
v_i^{G.I.}=\partial_i\left(B-a\dot{E}\right)+\frac{\delta T^0_{\,\,\,\, i}}{a(\rho+p)}
\label{vGI}
\ee
where $\rho$ and $p$ are the background energy density and pressure and $\delta T^0_{\,\,\,\, i}$ is a perturbation to the energy-momentum tensor. For completeness let us also give the gauge invariant energy density perturbation which is most conveniently written in Fourier space
\be
\epsilon^{G.I.}=\frac{\delta\rho}{\rho}+3H\frac{ik^i \delta T^0_{\,\,\,\, i}}{k^2\rho}.
\label{eGI}
\ee
where $k^i$ is the comoving wave number. Since the sum of two gauge invariant variables is also gauge invariant, Eq.(\ref{phiAH}) and Eq.(\ref{vGI}) can be used to define a new gauge invariant variable
\ba
\zeta^{G.I.}&=&\Phi_H+aH\frac{-ik^iv_i^{G.I.}}{k^2}\\
\label{zGI}
\ea
The expression for $\zeta^{G.I.}$ can be simplified by using the first order Einstein equations, for a universe filled with a perfect fluid with equation of state $p=w\rho$ we have
\be
\zeta^{G.I.}=\frac{5+3w}{3+3w}\Phi_H+\frac{2}{3(1+w)}H^{-1}\dot{\Phi}_H.
\ee
It can be shown that on large scales in an era where $w$ is constant $\dot{\Phi}_H=0$ so the above reduces to
\be
\zeta^{G.I.}=\frac{5+3w}{3+3w}\Phi_H
\label{ztoPhi}
\ee
where $w$ is the equation of state of the dominant energy component in the universe (e.g. $w=1/3$ during radiation domination and $w=0$ during matter domination).

\subsection{Relating $\delta\phi$ to $\zeta$ and $\zeta$ to $\delta$}
The ADM formalism (which is reviewed nicely in \cite{Misner:1974qy}) is useful for obtaining the relation between perturbations in the inflaton $\delta\phi$ and the gauge invariant scalar $\zeta^{G.I.}$. Here we follow the notation of \cite{Maldacena:2002vr} writing the metric with scalar perturbations as
 \ba
ds^2&=&-(1+N_1)^2dt^2+a^2(1+2\zeta)\delta_{ij}(dx^i+N^idt)(dx^j+N^jdt)\nn
N_i&=&a^2\delta_{ij}N^j\equiv a^2\partial_i\psi\;.
\label{MaldacenaMetric}
\ea
The gauge has not yet been fully specified, but we can fix it by requiring that $\zeta=0$ (spatially flat gauge) or by choosing slices with $T^0_{\,\,\,\, i}\propto \delta\phi=0$ (comoving gauge). Once the gauge is fixed the constraint equations can be used to eliminate $N_1$ and $\psi$ in favor of $\zeta$ or $\delta\phi$. Calculations in the two gauges can be related by using gauge invariant variables. 
With this parameterization of the metric 
\ba
\zeta^{G.I.}&=&\zeta-\frac{Hik^i\delta T^0_{\,\,\,\, i}}{k^2(E+P)}.
\label{zdef}
\ea
We can see that in the comoving gauge where $\delta T^0_{\,\,\,\, i}=0$, $\zeta^{G.I.}=\zeta$. While in the spatially flat gauge where $\zeta=0$ and $\delta T^0_{\,\,\,\, i}=-\dot{\phi}\partial_i\delta\phi$, $\zeta^{G.I.}=-H\delta\phi/\dot{\phi}$. Here we need to relate $\zeta$ to $\Phi_H$ during the matter dominated era, so  from Eq.(\ref{ztoPhi}), $\Phi_H=3/5\zeta$.
In conformal Newtonian gauge during the matter dominated era the Bardeen curvature $\Phi_H$ satisfies Poisson's equation
\be
k^2\Phi_H =4\pi G a^2\delta \rho.
\label{poisson}
\ee
where we are using the Fourier convention
\be
\delta(\vx)=\int \frac{d^3 \vk}{(2\pi)^3} e^{i\vk\cdot\vx}\delta(\vk).
\ee
Now, define the linear growth factor by $\Phi_H(\vk,z)=D(z)(1+z)\Phi_H(\vk)$. Putting this together with Eq.(\ref{ztoPhi}) gives the relation between $\delta$ and $\zeta$ in the matter dominated era as 
\be
\delta(\vk,z)=\frac{2}{5}\frac{1}{\Omega_m}\frac{1}{H_0^2}D(z)T(k)k^2\zeta(\vk).
\ee

\subsection{$f_{NL}$ Conventions}
\label{fNLconv}
In the literature there are several conventions for the definition of the non-Gaussian parameter $f_{NL}$. In this paper, we use the convention of \cite{Komatsu:2001rj} who define $f_{NL}$ in terms of the Bardeen curvature $\Phi_H$, 
\be
\Phi_H=\Phi_{H,G}+f_{NL}(\Phi_{H,G}^2-\VEV{\Phi_{H,G}^2})
\label{phiHGfNL}
\ee
where the extra subscript $G$ is to indicate that the field is a Gaussian random field. Physically, if $f_{NL}$ is defined as above in terms of the curvature, then a positive $f_{NL}$ leads to negative skewness in the temperature field and positive skewness in the density field (corresponding to \emph{more} rare objects).  

In a gauge where $B-a\dot{E}=0$ (for example in the often-used conformal Newtonian gauge), $\Phi_A=-\Phi_H$ \footnote{This statement is only true if anisotropic stress is unimportant.}. So if one instead used the above expansion in terms of the gravitational potential $\Phi_A$ then a minus sign would need to be introduced to compare with the WMAP $f_{NL}$. To be completely explicit, suppose we defined $\tilde{f}_{NL}$ by 
\be
\Phi_A=\Phi_{A,G}+\tilde{f}_{NL}(\Phi_{A,G}^2-\VEV{\Phi_{A,G}^2})
\label{phiAGfNL}
\ee
then comparing with Eq.(\ref{phiHGfNL}) we see $\tilde{f}_{NL}=-f_{NL}$. 

This sign change is indeed necessary to compare $f_{NL}$ here with $f_{NL}$ as given by Maldacena  \cite{Maldacena:2002vr} and CHKS \cite{Chen:2006nt}. Maldacena defines $f^{M}_{NL}$ by
\be
\zeta=\zeta_L-\frac{3}{5}f^{M}_{NL}\zeta_L^2
\ee
The $3/5$ appears because during the matter dominated era $\zeta=5/3\Phi_H=-5/3\Phi_A$.  In summary, 
\be
f_{NL}^{here} = +f_{NL}^{WMAP} = -f_{NL}^{M} 
\ee

Our definitions of $f_{NL}$ for sound speed models actually differ from those in CHKS \cite{Chen:2006nt} in two ways, first we have allowed for tilted spectrum $n_s\neq 1$ giving rise to the factors of $3^{n_s-1}$. Second, we differ by a minus sign because we have defined $f_{NL}$ here to agree with the WMAP convention -- i.e. $f_{NL}$ here, Eq.(\ref{phiHGfNL}), is defined in terms of the Bardeen curvature perturbation $\Phi_H$, while positive $f_{NL}$ in CHKS is defined in terms of the Newtonian potential $\Phi_A$.

The above discussion is consistent with a special case that can be understood physically. In slow roll, in the ``squeezed triangle" limit where $k_1\ll k_2, k_3$, the non-Gaussianity is proportional to the tilt of the scalar spectral index $n_s-1$. To see this, consider a smooth potential with $\partial_{\phi}V<0$ and relate the Hubble parameter and the variation of the inflaton $\phi$ by:
\be
\frac{H}{\dot{\phi}}\Delta\phi= H\Delta t
\ee
where $t$ is time and the dot denotes a time derivative. Since $H>0$, $\dot{\phi}>0$, a fluctuation $\Delta\phi<0$ (the field jumps up the potential) means $\Delta t<0$. This is the usual way of thinking of fluctuations of the inflaton field meaning that inflation ends at slightly different times in different patches. A fluctuation $\Delta\phi<0$ means that inflation ends a little later in that patch and so the patch is a hotter, higher density region. In the spatially flat gauge, where the curvature perturbation $\zeta$ (which appears in the metric) is zero, the gauge invariant curvature is related to the fluctuations of the inflaton field by $\zeta^{G.I.}=-H\delta\phi/\dot{\phi}$ (consistent with Maldacena's Eq.(2.26)). 

For the squeezed limit, mode $k_1$ is frozen out much earlier than $k_2$, $k_3$, so a fluctuation $\zeta>0$ only causes $k_2$, $k_3$ to exit the horizon a little earlier ($\Delta t=-\zeta/H$, exiting farther from the end of inflation in that patch, which occurs at the same value of $\phi$ as in all other patches). If the spectrum is blue-tilted, this gives negative skewness since there will be less power in the fluctuations in that patch, and so less power in the largest density fluctuations, than if the power spectrum was flat. In his Eq.(4.9), Maldacena finds $-f_{NL}^{M}\sim-(n_s-1)$. Then a blue tilt corresponds to $f_{NL}^{M}>0$ and negative skewness. We are using the opposite sign convention from Maldacena. In DBI, then, the largest effect will give negative skewness. In other models where the first term in Eq.(\ref{fnls}) dominates, the largest effect could give positive skewness. 

\section{The Probability Distribution Function}
\label{Distributions}

\subsection{Comparing the Various Distributions}
The $f_{NL}$ expansion is useful for examining the effect of keeping only a few terms in Edgeworth expansion\footnote{One can make a similar comparison with other expansions, such as the Gamma expansion developed in \cite{Gaztanaga:1999er}. The Gamma expansion is most useful for the case of positive skewness, so we have not developed it here.}. It is also useful since one can obtain the PDF by making a formal change of variable in the Gaussian distribution. Doing so gives
\ba
\label{ngdist}
P(\nu)d\nu&=&\frac{d\nu}{\sqrt{1+4f_{NL}\sigma(\nu+f_{NL}\sigma)}}\frac{1}{\sqrt{2\pi}}(\mathcal{E}_{-}+\mathcal{E}_{+})\\\nonumber
\mathcal{E}_{\pm}&=&\exp\left[-\frac{1}{4\sigma^2f_{NL}^2}\left(1+2f_{NL}\sigma(\nu+f_{NL}\sigma)\mp\sqrt{1+4f_{NL}\sigma(\nu+f_{NL}\sigma)}\right)\right]
\ea
Note that the natural combination here is $|f_{NL}\sigma|$. For $|f_{NL}\sigma|<1/10$, this distribution is smooth and resembles the Gaussian. At larger values the shape develops a feature on the suppressed tail. It also has an imaginary part on one tail for $f_{NL}<0$. This formal change of variable is useful for comparing various truncations of the Edgeworth expansion using unsmoothed cumulants. 

\subsection{Validity of the Edgeworth Expansion}
\begin{figure}[htb]
\begin{center}
$\begin{array}{cc}
\includegraphics[width=0.5\textwidth,angle=0]{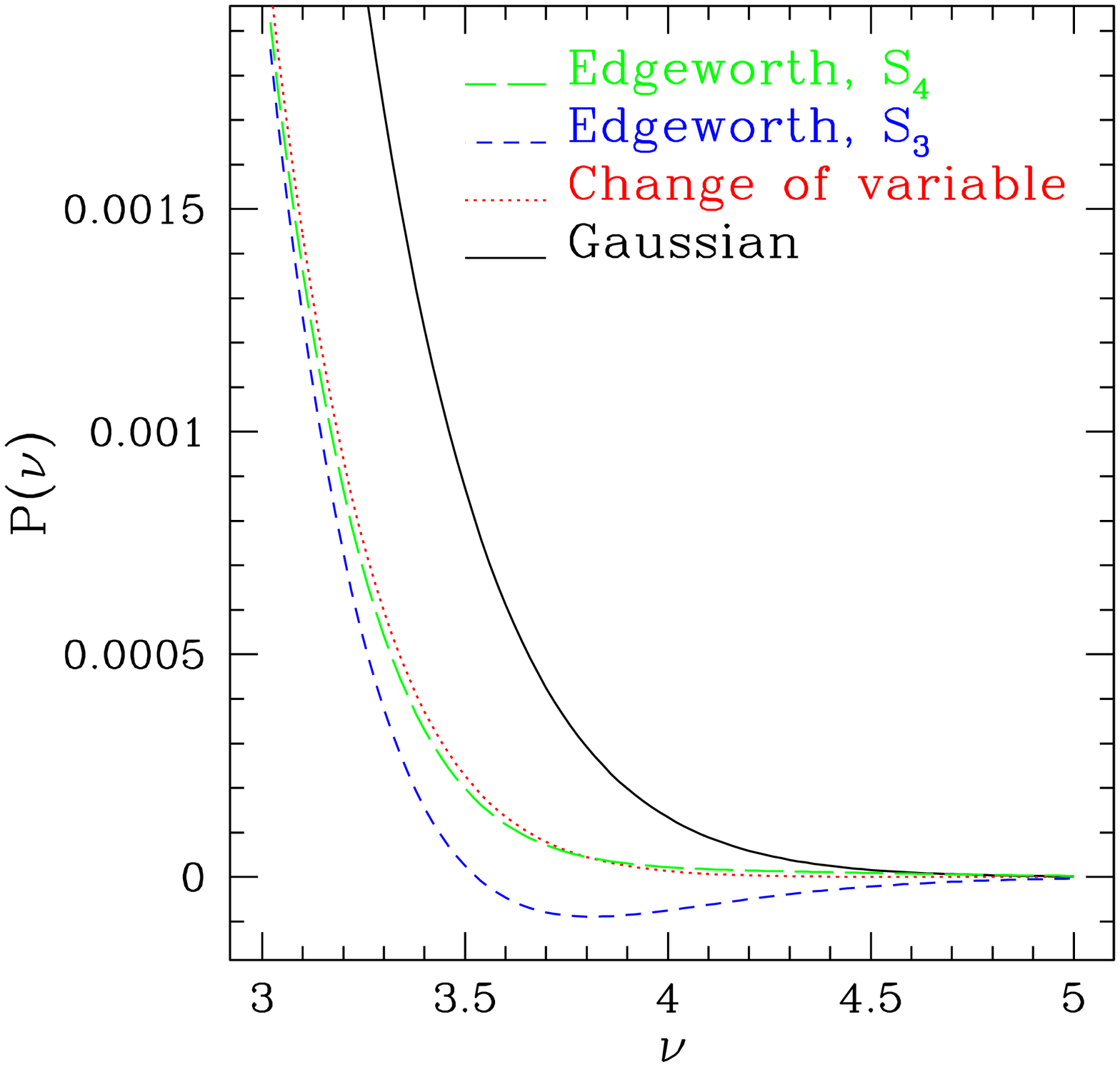} &
\includegraphics[width=0.5\textwidth,angle=0]{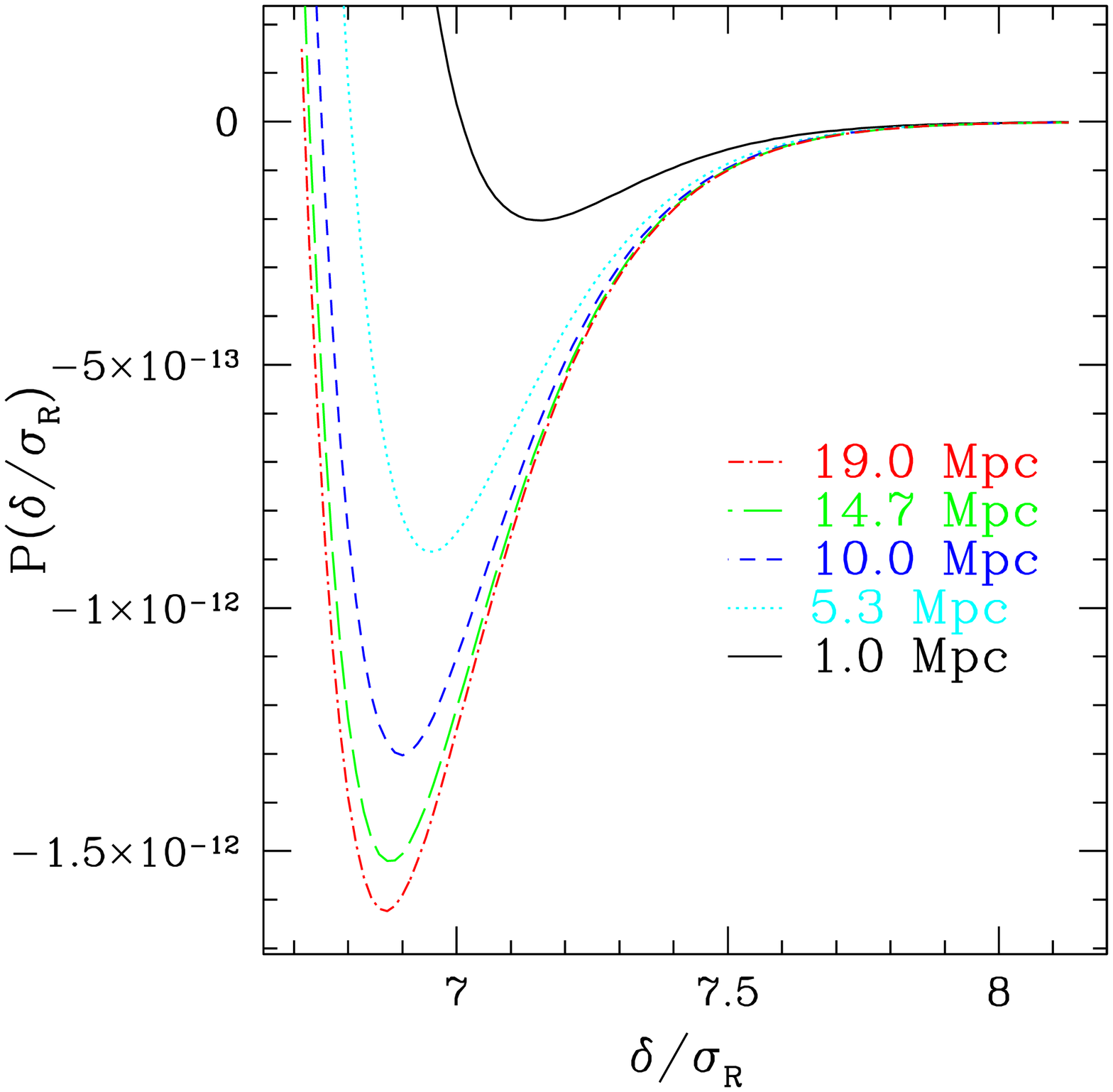} \\
\mbox{(a)} &\mbox{(b)}
\end{array}$
\caption{(a) A region of the probability distribution that is negative if the Edgeworth series is truncated at $S_3$. The distributions are plotted for unsmoothed quantities, with variance $\sigma=1$ and $f_{NL}=-0.03$. Notice that in this case, including the kurtosis cures the negative region. (b) Examination of the negative region of the PDF, coming from truncation of the Edgeworth expansion for $f_{NL}=-256$ and $\kappa=0$. Results for several values of the smoothing scale are shown. }
\label{negregions}
\end{center}
\end{figure}

The Edgeworth series, Eq.(\ref{Edgeworth}) is an exact expression for the PDF in terms of its cumulants, but in practice one often calculates just the first few cumulants. How valid is the PDF if we use, say, only the skewness? There are several issues here. First, assuming we keep only the first and second terms in the series Eq.(\ref{Ssum}) (up to linear in $S_3$), the steepest descent calculation is only valid for $\nu<1/|S_3\sigma|$. In addition, the final distribution will not be positive definite unless higher-order cumulants are included. We truncate the Edgeworth expansion at linear order in $S_3$, since $S_4$ comes in at roughly the same order as $S_3^2$. Panel (a) of Figure \ref{negregions} shows a comparison of a negative region for the change of variable distribution Eq.(\ref{ngdist}), the Edgeworth expansion, Eq.(\ref{Edgeworth}), including up to the $S_3\sigma$ term and the Edgeworth expansion including the $\sigma^2$ term. The Gaussian is also plotted for reference. With negative skewness, the distribution has a negative region on the $\nu>0$ side.

If $\delta/\sigma<1$ the Edgeworth expansion will be a good approximation to the true PDF so long as 
\be
1>> S_3\sigma>>S_4\sigma^2>>S_5\sigma^3 \dots
\ee
However for collapsed objects such as clusters $\delta/\sigma>1$. In this case the leading term in each Hermite polynomial (see Eq.(\ref{Edgeworth})) is $H_n\sim(\delta/\sigma)^n$. Then truncating the expansion requires something like
\be
S_3\sigma\left(\frac{\delta}{\sigma}\right)^3<<1\quad\textrm{ and }\quad S_n>> S_{n+1}\sigma\left(\frac{\delta}{\sigma}\right)\quad \textrm{ for }\quad n \ge 3
\ee
Now, the number of terms that should be included in the Edgeworth expansion depends on the value of $\delta/\sigma$. For clusters, $\delta/\sigma\sim (few)$ to $\delta/\sigma \rightarrow \infty$, we would in principle need an infinite number of the cumulants for a valid expansion. This is a point of concern, but since we will be interested in the PDF integrated above some threshold and the PDF drops off rapidly for $\delta/\sigma \gs 1$ the situation isn't as bad is it may appear. We expect then, to find an expression for the PDF that has a limited range of validity. The range will depend on mass, since $\sigma$ is a decreasing function of $M$, and on redshift since the threshold for a density fluctuation to collapse, $\delta_c$, is larger at high redshift. 

Below, panel (b) of Figure \ref{negregions} shows the Edgeworth expansion results for the DBI case ($\mathcal{A}_c$ term) with $f_{NL}=-256$ and $\kappa=0$. The growth of the negative region at larger scales corresponds with expectations from panel (d) of Figure \ref{deltasfig}.

\subsection{Validity of the Non-Gaussian Mass Function}
The derivation of the mass function in \S \ref{NGMF} involves integrating the PDF above the threshold for collapse $\delta_c$. Using the Edgeworth expansion and keeping terms up to second order one finds
\ba
P(>\delta_c,M)&=&\frac{1}{2}erfc\left[\frac{\delta_c}{\sqrt{2}\sigma_M}\right]+\frac{S_3(M)\sigma_M}{3!}\left(\frac{\delta_c^2}{\sigma_M^2}-1\right)\frac{e^{-\frac{\delta_c^2}{2\sigma_M^2}}}{\sqrt{2\pi}}\nn
&+&\frac{1}{2}\left( \frac{S_3(M)\sigma_M}{3!}\right)^2\frac{\delta_c}{\sigma_M}\left(\frac{\delta_c^4}{\sigma_M^4}-10\frac{\delta_c^2}{\sigma_M^2}+15\right)\frac{e^{-\frac{\delta_c^2}{2\sigma_M^2}}}{\sqrt{2\pi}}\nn
&+&\frac{S_4(M)\sigma_M^2}{4!}\frac{\delta_c}{\sigma_M}\left(\frac{\delta_c^2}{\sigma_M^2}-3\right)\frac{e^{-\frac{\delta_c^2}{2\sigma_M^2}}}{\sqrt{2\pi}}+\dots
\ea
From this a mass function with terms up to second order (compare with Eq.(\ref{massfcnNG}) which is to first order) can be obtained
\ba
\left.\frac{dn(M)}{dM}\right\vert_{2^{nd}}&=&-\sqrt{\frac{2}{\pi}}\frac{\bar{\rho}}{M}e^{-\frac{\delta^2_c}{2\sigma_M^2}}\left\lbrace\frac{d\textrm{ln}\sigma_M}{dM}\left[\frac{\delta_c}{\sigma_M}+\frac{S_3\sigma_M}{3!}\left(\frac{\delta_c^4}{\sigma_M^4}-2\frac{\delta_c^2}{\sigma_M^2}-1\right)\right.\right.\nn
&&+\left.\frac{1}{2}\left(\frac{S_3\sigma_M}{3!}\right)^2\left(\frac{\delta_c^7}{\sigma_M^7}-13\frac{\delta_c^5}{\sigma_M^5}+25\frac{\delta_c^3}{\sigma_M^3}+15\frac{\delta_c}{\sigma_M}\right)\right.\nn
&&\left.+\frac{S_4\sigma_M^2}{4!}\left(\frac{\delta_c^5}{\sigma_M^5}-4\frac{\delta_c^3}{\sigma_M^3}-3\frac{\delta_c}{\sigma_M}\right)\right]\nn
&&+\left.\frac{\sigma_M}{6}\frac{dS_3}{dM}\left(\left(\frac{\delta_c^2}{\sigma_M^2}-1\right)+\frac{S_3\sigma_M}{3!}\frac{\delta_c}{\sigma_M}\left(\frac{\delta_c^4}{\sigma_M^4}-10\frac{\delta_c^2}{\sigma_M^2}+15\right)\right)\right.\nn
&&+\left.\frac{\sigma_M^2}{4!}\frac{dS_4}{dM}\frac{\delta_c}{\sigma_M}\left(\frac{\delta_c^2}{\sigma_M^2}-3\right)\right\rbrace
\label{massfcnNG2order}
\ea

Without knowledge of all higher cumulants we are forced to use an approximate form for the mass function. Since the skewness is the dominant contributer, we keep terms up to $S_3\sigma$. We assume that when the $(S_3\sigma)^2$ terms become important, it is no longer valid to neglect the $S_4\sigma^2$ term. Similarly, we are assuming that the $S_4\sigma^2$ terms become important no sooner than the $(S_3\sigma)^2$ terms. Figure \ref{nofMcomp}  compares the mass functions Eq.(\ref{massfcnNG}) and Eq.(\ref{massfcnNG2order}) when one and two orders of $S_3\sigma$ are kept. We define the regime of validity for our truncated mass function (Eq.(\ref{massfcnNG})) to be where corrections from including the $(S_3\sigma)^2$ in Eq.(\ref{massfcnNG2order}) reach $5\%$. One can see that the range of mass scales where the mass function Eq.(\ref{massfcnNG}) is valid decreases both with redshift and as $f_{NL}$ increases in magnitude.

\subsection{Comparison with MVJ}
\label{AppMVJ}
\begin{figure}[!t]
\begin{center}
$\begin{array}{cc}
\includegraphics[width=0.5\textwidth,angle=0]{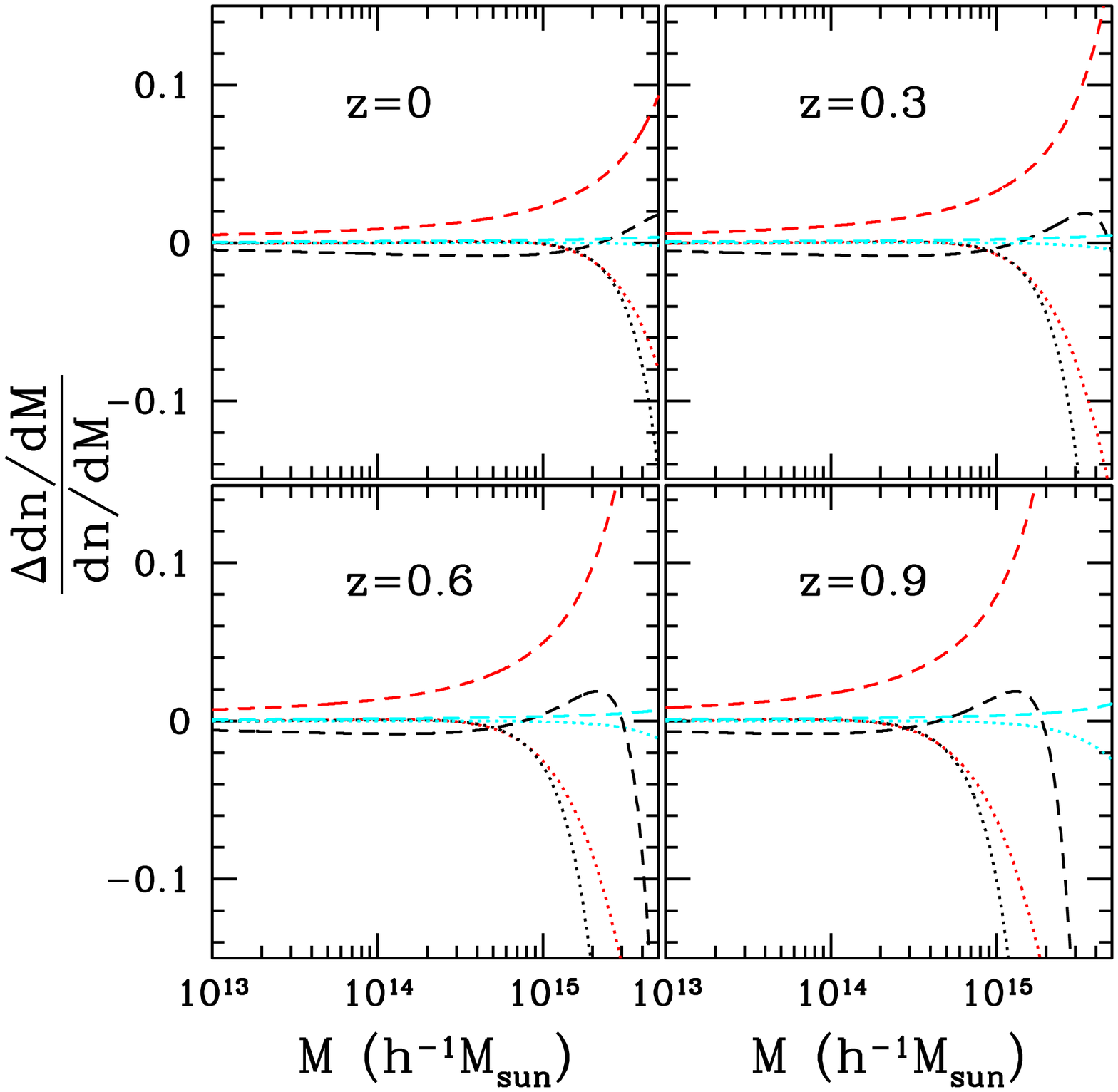} 
&\includegraphics[width=0.5\textwidth,angle=0]{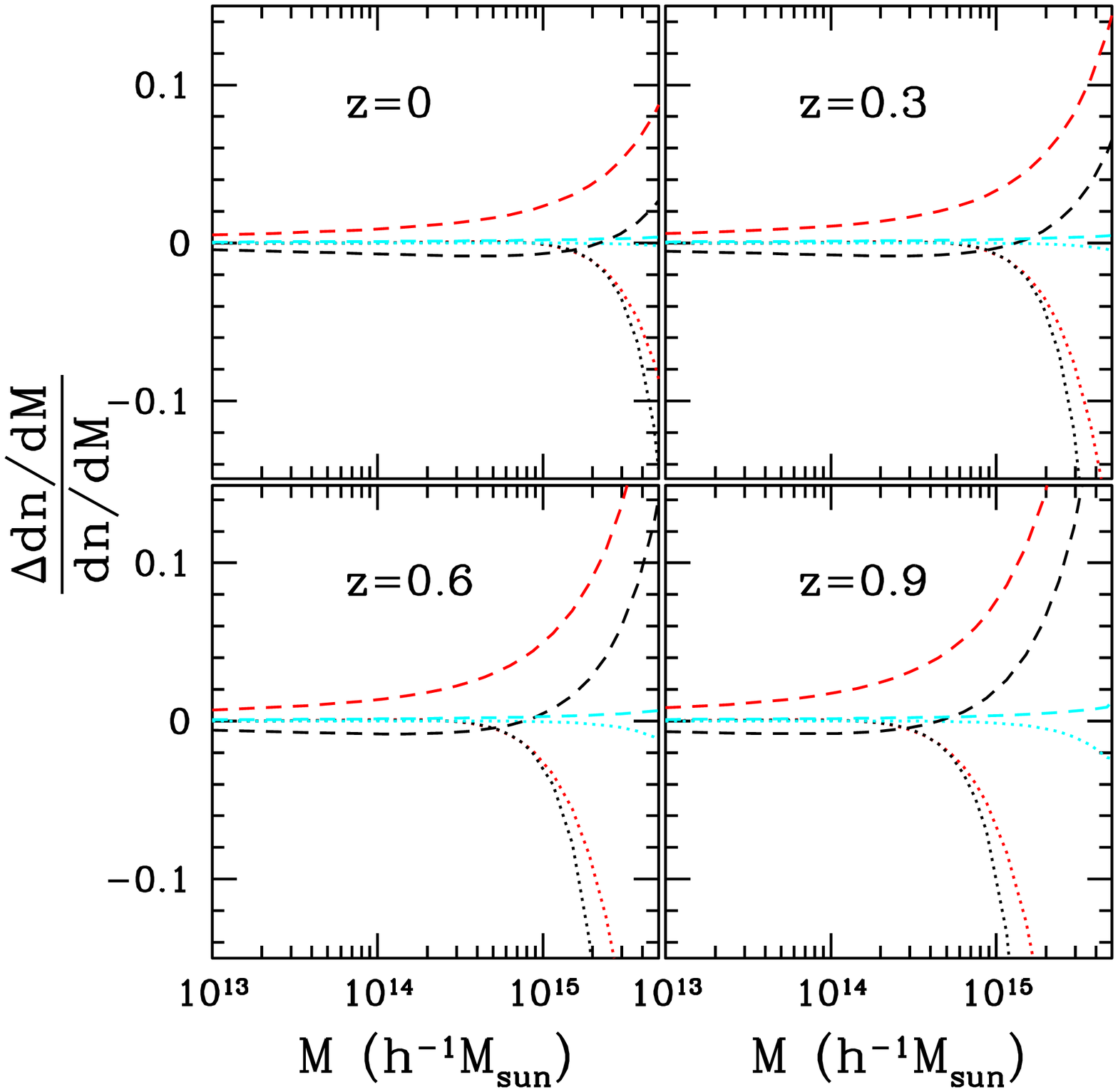} \\
\mbox{(a)} &\mbox{(b)}
\end{array}$
\caption{Comparison of the different mass functions for different non-Gaussian models, shown are  $f_{NL}^{eq}=-256$ (black), $332$ (red) and $38$ (cyan).(a) Dotted lines: Comparison of the first order in $S_3\sigma$ mass function Eq.(\protect\ref{massfcnNG}) and the second order in $S_3\sigma$ mass function Eq.(\protect\ref{massfcnNG2order}), plotted is  (Eq.(\protect\ref{massfcnNG})-Eq.(\protect\ref{massfcnNG2order}))/Eq.(\protect\ref{massfcnNG2order}) with $S_4=0$.  Dashed lines: Comparison of the MVJ mass function Eq.(\protect\ref{massfcnMVJ}) and the second order in $S_3\sigma$ mass function Eq.(\protect\ref{massfcnNG2order}), plotted is (Eq.(\protect\ref{massfcnMVJ})-Eq.(\protect\ref{massfcnNG2order}))/Eq.(\protect\ref{massfcnNG2order}) with $S_4=0$.  (b) Dotted lines: Comparison of the first order in $S_3\sigma$ mass function Eq.(\protect\ref{massfcnNG}) and the numerically integrated all-orders in $S_3\sigma$ mass function Eq.(\protect\ref{massfcnallorder}), plotted is  (Eq.(\protect\ref{massfcnNG})-Eq.(\protect\ref{massfcnallorder}))/Eq.(\protect\ref{massfcnallorder}).  Dashed lines: Comparison of the MVJ mass function,  Eq.(\protect\ref{massfcnMVJ})  and the all-orders in $S_3\sigma$ mass function Eq.(\protect\ref{massfcnallorder}), plotted is (Eq.(\protect\ref{massfcnMVJ})-Eq.(\protect\ref{massfcnallorder}))/Eq.(\protect\ref{massfcnallorder})
}
\label{nofMcomp}
\end{center}
\end{figure}

In \S \ref{NGMF} we presented a mass function using the Edgeworth expansion for the non-Gaussian probability distribution. Another form for the non-Gaussian mass function, based only on the skewness was given in Matarrese, Verde and Jimenez \footnote{Here we correct a typo in Eq. (68) of MVJ, this expression is in agreement with Eq. (6) of \cite{Verde:2000vr}.} (hereafter MVJ) \cite{Matarrese:2000iz}. 
\be
\frac{dn}{dM}_{MVJ}(M,z)=-\frac{2\bar{\rho}}{M^2}\frac{1}{\sqrt{2\pi}\sigma_M}\left[\frac{1}{6}\frac{\delta_c^3}{\delta_*}\frac{dS_3}{d\ln M}+\delta_*\frac{d\ln \sigma_M}{d\ln M}\right]e^{-\delta_*^2/(2\sigma_M^2)}
\label{massfcnMVJ}
\ee
where $\delta_*=\delta_c\sqrt{1-S_3\delta_c/3}$. Neither Eq.(\ref{massfcnNG}) nor Eq.(\ref{massfcnMVJ}) are exact expressions for the number density of massive objects. There are different ways to test these approximations. Perhaps the most reliable method is to compare with N-body simulations. Figure 3 of \cite{Grossi:2007ry} shows that for values of $f^{local}_{NL} \sim 100$ and for reasonable values of $M$  (at least the range contained in Figure \ref{nofMcomp}) and $z=0$ the MVJ approximation is in good agreement with simulations. However a direct comparison of mass functions from different simulations \cite{Kang:2007gs, Grossi:2007ry, Dalal:2007cu}  should be done before drawing definitive conclusions. A more detailed comparison with simulations (including for example the equilateral shape non-Gaussianity) is left to future work. 

Short of running simulations, one way to compare the approximations is, as discussed in the previous section, to consider corrections from including the $(S_3\sigma)^2$. In Figure \ref{nofMcomp} we plot the fractional difference between the mass function of MVJ, Eq.(\ref{massfcnMVJ}) and Eq.(\ref{massfcnNG2order}). While for $f_{NL}=332$ Eq.(\ref{massfcnNG}) is a better approximation to the second order mass function Eq.(\ref{massfcnNG2order}) than MVJ, for $f_{NL}=38$ and $-256$ the mass function of MVJ is a better approximation. 

Another comparison one may make is to assume the (possibly unphysical) case of non-Gaussianity that produces only $S_3$, with all higher cumulants vanishing. In this case, an expression for the mass function can be determined from Eq.({\ref{cumulantPDF}) and Eq.(\ref{dndmdef}) where $S(y)$ in Eq.  ({\ref{cumulantPDF}) is truncated at $S_3$. Exchanging the orders of integration ($d\delta \leftrightarrow dy$) one arrives at MVJ Eq. (62) from which we can define a mass function that keeps all orders in $S_3\sigma$
\be
\frac{dn}{dM}_{\textrm{{\tiny all orders}}}(M,z)=\frac{2}{\pi}\frac{\bar{\rho}}{M}\frac{d}{dM}\left[\int_0^\infty \frac{d \lambda}{\lambda} e^{-\lambda^2\sigma_M^2/2}\sin\left(\lambda\delta_c(z)+\frac{S_3(M)\sigma_M^4\lambda^3}{6}\right)\right]
\label{massfcnallorder}
\ee
where $\lambda=iy/\sigma_M^2$. The above expression must be numerically integrated. A comparison of the all-orders mass function with the first order and MVJ mass functions is shown in panel (b) of Figure \ref{nofMcomp}.  For $f_{NL}=38$ and $-256$ the mass function of MVJ more closely matches the all-orders mass function.  While for $f_{NL}=332$ Eq.(\ref{massfcnNG}) is in better agreement.

In this paper we have used the mass function derived from the Edgeworth expansion because we suspect that higher cumulants (e.g. $S_4$) may become important at high mass and/or high redshift. With the Edgeworth-derived mass function we have a better analytic understanding the range of masses and redshifts where we expect the mass function to be valid when truncated at $S_3$.

\bibliographystyle{JHEP}

\end{document}